\documentclass[aps,pra,twocolumn,groupedaddress]{revtex4-2}

\usepackage{graphicx}
\usepackage{siunitx}
\usepackage{color}
\usepackage{amsmath}

\begin{document}


\title{Superfluid $^4$He as a rigorous test bench for different damping models in nanoelectromechanical resonators}


\author{T. Kamppinen}
\email[]{timo.kamppinen@aalto.fi}
\author{J. T. M{\"a}kinen}
\author{V. B. Eltsov}
\affiliation{Low Temperature Laboratory, Department of Applied Physics, Aalto University}


\date{\today}

\begin{abstract}
We have used nanoelectromechanical resonators to probe superfluid $^4$He at different temperature regimes, spanning over four orders of magnitude in damping. These regimes are characterized by the mechanisms which provide the dominant contributions to damping and the shift of the resonance frequency: tunneling two level systems at the lowest temperatures, ballistic phonons and rotons at few hundred \SI{}{\milli \kelvin}, and laminar drag in the two-fluid regime below the superfluid transition temperature as well as in the normal fluid. 
Immersing the nanoelectromechanical resonators in fluid increases their effective mass substantially, decreasing their resonance frequency.
Dissipationless superflow gives rise to a unique possibility to dramatically change the mechanical resonance frequency in situ, allowing rigorous tests on different damping models in mechanical resonators.
We apply this method to characterize tunneling two-level system losses and magnetomotive damping in the devices.
\end{abstract}

\maketitle

\section{Introduction}

Nanoelectromechanical (NEMS) resonators have emerged in many fields of physics as ultra-sensitive probes of mass and force \cite{LiMo2007}. For instance, their extreme force resolution enabled measurements sensitive to magnetic field of a single nuclear spin \cite{Myller2014}. 
Recently, detection of a single quantized vortex trapped on a NEMS device in superfluid $^4$He has been demonstrated \cite{Guthrie2021}. 
Trapping of a single vortex on a NEMS device in $^3$He will open new avenues for exciting studies, for example on Majorana zero modes living in quantized vortex cores \cite{Kopnin1991}, the building blocks of a topologically protected quantum computer.

Full understanding of intrinsic device properties and device-fluid interactions is required for detailed analysis of high-precision measurements e.g. on vortex dynamics, and superfluid $^4$He is an excellent sandbox for studying those.
In this work we provide detailed description of device properties and device-fluid interactions for NEMS resonators of different sizes immersed to the superfluid $^4$He, and analyze our results using existing theoretical models.

We compare the response of the same devices in vacuum and in superfluid, which provides additional information about the intrinsic device properties.
Density of thermal excitations in superfluid $^4$He becomes vanishingly small at temperatures $T\lesssim \SI{0.2}{\kelvin}$, and in absence of quantized vortices, it is practically an ideal fluid with frictionless superflow.
The ideal flow of the liquid displaced by the device reduces the resonance frequency via an increase in the effective mass of the resonator, but it does not introduce extra damping. 
The ability to change both temperature and frequency of a mechanical resonance mode in the same device allows rigorous tests on different damping models. We apply these tests in particular to tunneling two level systems (TTLS) and magnetomotive damping mechanisms observed in our devices \cite{Kamppinen2022}. 
Beyond nanoelectromechanical resonators, TTLS affect damping, noise and decoherence in a wide range of quantum-limited measurements \cite{Behunin2016}, for example in qubits \cite{Ithier2005} and in optomechanical systems \cite{Laure2021}. 
As the dimensions of those devices are often smaller than the relevant phonon wavelengths, interest in TTLS properties in reduced dimensions extends beyond simple mechanical resonators \cite{Behunin2016}.

Our experiments reveal that the different damping mechanisms scale differently with mass and frequency. 
In particular, damping rate $\Delta f$ due to magnetomotive damping is found to scale with the resonance frequency $f_0$ as $ \Delta f \propto f_0^2 \propto 1/m $ due to increase in the effective mass $m$ of the device.
Damping rate resulting from TTLS is found to scale approximately as $\Delta f \propto f_0 \propto 1/(m f_0) $ in the low-temperature regime where TTLS relaxation rate is slow compared to the device frequency. 
Such scaling is not expected from the currently adopted extensions of the standard tunneling model to reduced dimensions \cite{Behunin2016,Seoanez2008,Kamppinen2022} without changes in the phonon-TTLS coupling parameter $\gamma$. 
We attribute the observed scaling to change in the coupling parameter $\gamma_1$ describing TTLS interaction with the strain field at the device frequency, while the coupling parameter $\gamma_2$ describing the TTLS relaxation via a pool of phonons is expected to remain unchanged. In our devices $\gamma_1$ scales approximately as $\gamma_1 \propto f_0^{-1/2}$.
The proposed scalings with mass and frequency can be used as an aid when assessing damping in geometries where current theoretical models are not directly applicable. 

A plethora of mechanical devices, including quartz tuning forks \cite{Blaauwgeers2007,Bradley2009}, vibrating wires \cite{Yano2005}, grids \cite{Efimov2009}, spheres \cite{Jager1995}, and recently MEMS and NEMS devices \cite{Defoort2016,Guenault2019,Zheng2016} have been used to probe the properties of the quantum fluids $^3$He and $^4$He. 
The pursuit for higher sensitivity has led to dramatic reduction in the size of the devices, and an increase in the surface-area to volume ratio. Thus, the mass enhancement from the fluid is becoming more and more important. 
We hope that our systematic approach to account for the mass enhancement in the analysis will make comparison of results obtained with different types of devices more straightforward.

Density of thermal excitations in superfluid $^4$He increases with increasing temperature, and at $T \gtrsim \SI{0.2}{\kelvin}$, momentum exchange with ballistic quasiparticles takes over as the dominant dissipation mechanism in our devices. At $T \gtrsim \SI{0.7}{\kelvin}$, interactions between quasiparticles become important, and viscous drag dominates dissipation. We explain the devices' response using existing models in all the temperature regimes, and achieve good understanding of device-fluid interactions. This is a prequisite for detailed analysis of vortex dynamics in superfluid $^4$He, and in superfluid $^3$He where the physics is more involved. Remarkably, the devices are useful thermometers in all the temperature regimes, spanning over four orders of magnitude in dissipation.

\section{Methods}

\begin{figure}[b]
\includegraphics[width=0.9\columnwidth]{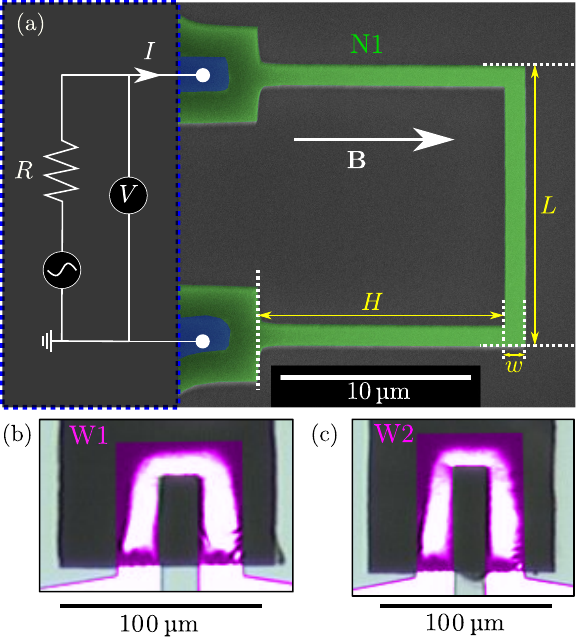}
\caption{(a) False-color electron micrograph of sample N1, and outline of magnetomotive measurement scheme. The dimensions ($H$,$L$,$w$) are tabulated in Table~\ref{table:dimensions}. The beam thickness $d$ is normal to the page.
The thin beam sections are freestanding after isotropic HF-vapor etch, while the wider sections are anchored on top of the substrate (blue regions). 
The sample is installed in a dilution refrigerator operated at a low temperature, while the measurement electronics is at room temperature. The motion of the device, $\dot{x}$, is directed in/out of the page. (b-c) False-color optical micrographs of the samples W1 and W2. The devices (magenta) are suspended over an orifice (black background) in the silicon substrate.}
\label{fig:sample}
\end{figure}

Suspended $\Pi$-shaped aluminum NEMS devices, comprising of two rectangular legs, and a rectangular paddle connecting the two legs, have been nanofabricated. Three different devices, shown in Fig.~\ref{fig:sample}, have been studied in this work.
The fabrication process and characterization of the devices in vacuum are described in Refs. \cite{Kamppinen2019,Kamppinen2022} and the dimensions of the devices are tabulated in Table~\ref{table:dimensions}.

\begin{table}[b]
	\centering
	\caption{The dimensions of the NEMS devices studied in this work. The dimension $L,H,w$ are taken from electron micrographs such as shown in Fig.~\ref{fig:sample}. The thickness $d$ is given by quartz crystal microbalance installed in the electron beam evaporator.}
\label{table:dimensions}
\begin{ruledtabular}	
\begin{tabular}{ r | r r r } 
Device 										& N1 				& W1 				& W2  \\
$L$  (\SI{}{\micro \meter})   				& 14.7       		& 60					& 60 			\\
$H$  (\SI{}{\micro \meter})   				& 13.0       		& 44   				& 60 			\\
$w$  (\SI{}{\micro \meter})   				& 1.11 	        		& 20   				& 20 			\\
$d$  (\SI{}{\nano \meter})    				& $150 \pm 8$  		& $200 \pm 10$  		& $200 \pm 10$ 	\\
\end{tabular}
\end{ruledtabular}
\end{table}

Device motion is excited and measured magnetomotively, as shown in Fig.~\ref{fig:sample}. 
The motion of the device  is driven with the Laplace force $ F = I L B $, where $L$ is the length of the beam perpendicular to the magnetic field~$\bf{B}$, and $I$ is the AC excitation current, produced by an arbitrary waveform generator connected in series with a resistor and the device. The motion of the device, $\dot{x}$, induces voltage, $ V = \dot{x} L B $ which is measured with a lock-in amplifier that is phase-locked with the generator.
Here $x(t)$ is the displacement of the device paddle from the equilibrium position with the maximum amplitude $x_0$.
The velocity and displacement amplitude are related via $v_0 = 2 \pi f {x_0}$, where $f$ is the frequency in \SI{}{\hertz}.
We deduce in-phase and quadrature displacement amplitudes $x_c$ and $x_s$, respectively, from the phase-resolved measurement.

\begin{figure}[t]
\includegraphics[width=\columnwidth]{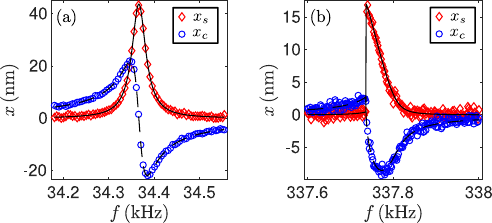}
\caption{Example spectra measured in superfluid $^4$He at \mbox{$T=\SI{20}{\milli \kelvin}$}. Solid and dashed lines are fits to Eqs.~(\ref{eq:xsin_mod_lorentz}) and (\ref{eq:xcos_mod_lorentz}), respectively. (a) Device W1, linear response. We set $D=0$ in the fitting. (b) Device N1, nonlinear response. Fitting is done with the full equations, giving $D = \SI{ -0.19 }{\hertz / nm^2}$.}
\label{fig:example}
\end{figure}

The equation of motion relevant for our NEMS resonators is \cite{Collin2010,Kamppinen2022} 
\begin{equation} \label{eq:dynamics}
\begin{split}
m( 1 + m_1 x + m_2 x^2) \ddot{x} + m (m_1/2 + m_2 x) \dot{x}^2  \\ +  2 \pi \Delta f m \dot{x}  + k(1 + k_1 x + k_2 x^2) x = F_0 \cos(2 \pi f t) ,
\end{split}
\end{equation}
where $m$ and $k$ are the (linear) effective mass and spring constant, $\Delta f$ is the damping rate (in \SI{}{\hertz}), $m_i$ and $k_i$ are nonlinear coefficients of mass and spring constant, respectively, and $F_0$ is the amplitude of the driving force. 
The wide devices W1 and W2 are always operated in the linear regime, and the nonlinear terms are irrelevant for these devices. The narrow device N1 becomes nonlinear at the lowest temperatures in vacuum and in fluid. For the small oscillation amplitudes considered in this work, the damping rate $\Delta f$ does not depend on the oscillation amplitude, but the resonance frequency changes due to the nonlinear terms.
We fit the response as a function of frequency to a modified Lorentzian, taking into account the nonlinear frequency shift \cite{Collin2010}
\begin{equation} \label{eq:xsin_mod_lorentz}
x_s(f) =  \frac{F_0}{m}  \left( \frac{1}{2\pi} \right)^2 \frac{ (\Delta f) f}{(f_r^2 - f^2)^2 + (\Delta f^2) f^2}
\end{equation}
and
\begin{equation} \label{eq:xcos_mod_lorentz}
x_c(f) = \frac{F_0}{m}  \left( \frac{1}{2\pi} \right)^2 \frac{f_r^2 - f^2}{(f_r^2 - f^2)^2 + (\Delta f^2) f^2},
\end{equation}
where the resonance position $f_r$ is a function of the squared displacement amplitude $ x_0^2 = (x_1^s)^2 + (x_1^c)^2 $ 
\begin{equation} \label{eq:freq_pulling}
f_r = \sqrt{ f_0^2 + 2 f_0 D x_0^2 } \approx f_0 + D x_0^2,
\end{equation}
where $D$ is, to a first approximation, a constant. For a linear response $D=0$.
We extract from the fits the resonance frequency $f_0$ and the damping rate $\Delta f$ of the mechanical resonator.
The linear resonance frequency is given by 
\begin{equation} \label{eq:f0}
f_0 = \frac{1}{2 \pi} \sqrt{\frac{k}{m}}.
\end{equation}
Example responses are shown in Fig.~\ref{fig:example}.
In the experiments we study how the response is affected by the temperature, magnetic field, and the presence of superfluid $^4$He.
Especially changes in the damping rate $\Delta f$ and resonance frequency $f_0$ extracted from the fits are particularly useful, as they give direct information about dissipative and reactive forces acting on the device.

The devices are installed in a hermetically sealed container, which is attached to a mixing chamber stage of a dilution refrigerator. The temperature of the mixing chamber stage is monitored with a ruthenium oxide thermometer and controlled with a resistive heater mounted on the mixing chamber plate.
The container is connected via a thin capillary to a gas handling system at room temperature. 
For vacuum measurements, the container is first flushed at room temperature with helium gas, and then pumped with a turbomolecular pump for at least 10 hours before cooling down the cryostat. 
The container has on its bottom a silver sinter with area $\sim \SI{10}{\m^2}$ for thermalizing the fluid. In vacuum measurements, the sinter acts as a cryopump for residual $^4$He gas.
For fluid measurements, helium gas with a nominal purity of 6.0 (impurity content $<1\,{\rm ppm}$) is admitted to the container slowly via a capillary filling line while the container is kept at a low temperature ($T<\SI{1}{\kelvin}$). This filling procedure is expected to limit amount of remnant vortices in the experimental cell \cite{Yano2005}.

The sample N1 was measured at \SI{2.6}{\bar} fluid pressure and the the samples W1 and W2 were measured at \SI{3.0}{\bar} fluid pressure, taken at $T=\SI{20}{\milli \kelvin}$. The volume of fluid in the capillary filling line is negligible compared to the volume in the experimental cell, and while the pressure changes as a function of temperature, fluid density is assumed to be constant.
The fluid density and velocity of sound is obtained from Ref. \cite{Abraham1970}, using for the zero-pressure reference density the value given in Ref. \cite{Donnelly1998}.
For roton properties, we use results of recent high-resolution neutron scattering experiments, Ref. \cite{Godfrin2021}.
The normal fluid ratio at elevated pressures is obtained by interpolating the tabular data in Refs. \cite{Brooks1977,Donnelly1998}, setting the normal fluid ratio to unity at the lambda-transition temperature. The temperature of the lambda transition at elevated pressures is obtained from Ref. \cite{Carty1973}.
For the viscosity, we use the values at saturated vapor pressure, Ref. \cite{Donnelly1998}, where we scale the viscosity with the fluid density used in our experiments, and the temperatures with the superfluid transition temperature corresponding to the pressure in the experiment.
 
\section{Theory} \label{section:theory}

Damping rate of a NEMS device has contributions from different mechanisms, such as tunneling two-level systems damping, magnetomotive damping, damping due to ballistic phonons and rotons, hydrodynamic drag, and possible other contributions, like temperature-independent clamping losses. We assume that the different contributions to the damping rate are additive.

In our experiments, we change the effective mass, $m$, of the resonator substantially by immersing the device in superfluid $^4$He. 
The expected effective mass enhancement in the fluid is \cite{Blaauwgeers2007}
\begin{equation} \label{eq:added_mass}
\left( \frac{ m_{\rm LHe}  }{ m_{\rm vac} } \right) = 1 + \beta \frac{\pi w}{4 d} \frac{ \rho_{\rm He} }{ \rho_{\rm Al} } +   \frac{B A}{\rho_{\rm Al} V} \sqrt{ \frac{\rho_n \eta}{\pi f_0} },
\end{equation}
where $\rho_{\rm He}$ is the density of the fluid, $\rho_n$ is the density associated with only the normal component of the fluid, $\rho_{\rm Al}$ is the density of aluminum, $A$ and $V$ are the surface area and volume of the device, and $\beta$ and $B$ are numerical constants of the order of unity \cite{Sader1998}.
At low temperatures, $T\lesssim \SI{0.7}{\kelvin}$, density of quasiparticles is low, inter-quasiparticle interactions leading to viscosity are irrelevant, and the last term in the right-hand side of Eq.~(\ref{eq:added_mass}) can be neglected. 
Experimentally, the mass enhancement is determined from the ratio of resonance frequencies in vacuum and in fluid
\begin{equation} \label{eq:exp_mass_enhance}
\left( \frac{ m_{\rm LHe}  }{ m_{\rm vac} } \right) = \left( \frac{ f_{0,{\rm vac}} }{ f_{0,{\rm LHe}} } \right)^2.
\end{equation}
The proportionality is obtained from Eq. (\ref{eq:f0}), assuming $k$ remains constant. 
In our experiments, a few solid layers of $^4$He formed on the device surface, when immersed in superfluid $^4$He are expected to have negligible contribution to $k$ \cite{Fear2016}.

At the lowest temperatures $^4$He does not contribute to dissipation, which allows us to compare intrinsic damping mechanisms at different mass and frequency for the same mechanical mode.
Thus, care has to be taken to use proper scaling when converting the parameter $\Delta f$ obtained from the measurements to relevant physical quantities. 
Dissipative forces that are proportional to the velocity are generally called laminar drag, and the drag force is given by the term 
\begin{equation} \label{eq:drag_force}
F_d = 2 \pi \Delta f m \dot{x}
\end{equation}
in the dynamics equation (\ref{eq:dynamics}). The corresponding dissipated power is $P_d = F_d \dot{x} $.
Examples of laminar drag are viscous drag at low velocities, drag due to momentum transfer with ballistic quasiparticles, magnetomotive damping, and intrinsic TTLS damping in the devices. 
For drag forces explicitly proportional to the velocity, we have
\begin{equation} \label{eq:massscale}
\Delta f \propto 1/m \propto f_0^2
\end{equation}
for the same drag force $F_d$ or dissipated power $P_d$.
At high velocities, turbulent drag force proportional to the squared velocity is expected \cite{Vinen2014}.
We restrict this work to the laminar regime where $\Delta f$ is independent of the velocity. 

Quite often, damping is characterized via the inverse quality factor, $ Q^{-1} =  \Delta f / f_0 $, which relates the energy lost per cycle to the energy stored in the oscillations. 
For damping mechanisms explicitly proportional to the velocity, mass enhancement is expected to increase the intrinsic quality factor via the additional energy stored in the oscillations.

\subsection{Temperature-independent damping}

Temperature independent contributions to the damping have been identified in Ref. \cite{Kamppinen2022}. A significant contribution results from the magnetomotive damping
\begin{equation}\label{eq:dfmm}
\Delta f_m = a_m B^2,
\end{equation}
where $a_m$ is a device- and measuring-circuit-dependent parameter and $B$ is the magnetic field strength.
In \cite{Kamppinen2022}, an analytical expression for $a_m$ relevant to our device geometry is given:
\begin{equation} \label{eq:mm_estimate}
a_m =  \frac{L^2 d}{3 m \rho_e } \left( \frac{w}{H+w} \right)^2,
\end{equation}
where $\rho_e$ is the electrical resistivity of aluminum at low temperature. Magnetomotive damping is predicted to be independent of frequency, and should scale with the effective mass as $\Delta f_m \propto 1/m$.

In vacuum, the wide devices W1 and W2 demonstrate additional temperature-independent contribution to the damping \cite{Kamppinen2022}, possibly via acoustic emission to the substrate. 
When device is immersed in a fluid, acoustic emission to the fluid is possible as well. 
A dipole emission model for acoustic emission of NEMS devices is suggested in Ref. \cite{Guenault2019}, resulting in
\begin{equation} \label{eq:acoustic_loss}
Q_{\rm ac}^{-1} = \frac{\pi^3}{2} \frac{ \rho_{\rm He} }{ \rho_{\rm Al} } \left( \frac{d_{\rm eff} f_0}{ c_{p}^2 } \right)^2 ,
\end{equation}
where $d_{\rm eff} \approx w$ is the effective beam diameter, and $c_p$ is the speed of sound in the fluid \cite{Abraham1970}.
For our devices, this expression predicts $Q_{\rm ac}^{-1}$ in the range \mbox{$ 10^{-5} \-- 10^{-6} $}, which is negligible compared to other damping mechanisms. 

\subsection{TTLS damping} \label{section:TTLS_theory}

Our devices have been characterized in vacuum before immersing into $^4$He. It has been found that the main dissipation mechanism at low temperatures is TTLS damping \cite{Kamppinen2022}.
The thermal phonon wavelength, $\lambda_{\rm ph} = (hc)/(k_B T)$, exceeds transverse dimensions of NEMS devices at low temperatures, and the dominant contribution to TTLS relaxation occurs via flexural phonon modes \cite{Kamppinen2022,Behunin2016,Seoanez2008}.
In the wider devices W1 and W2, the wavelength $\lambda_{\rm ph}$ exceeds only the thickness $d$, making these devices quasi-2D devices. In the narrow device N1, the wavelength $\lambda_{\rm ph}$ exceeds also the beam width $w$, making it a quasi-1D device.

At low temperatures, TTLS relaxation rate is slow compared to the oscillation frequency of the device, and TTLS damping in 1D devices is given by \cite{Behunin2016,Kamppinen2022}
\begin{equation}\label{eq:dfTTLS1D}
\Delta f_{\rm TTLS,1D} \approx 0.30 \frac{P_0 \gamma^4}{E^2} \frac{1}{c^{1/2} w d^{3/2} } \frac{ (k_B T)^{1/2} }{ \hbar^{3/2}},
\end{equation}
where $P_0$ is the TTLS density of states, $\gamma$ is the TTLS-phonon coupling, $E$ is the Young's modulus, and \mbox{$c=\sqrt{E/\rho_{\rm Al}}$} is the speed of sound in the aluminum beams.
In 2D devices the TTLS damping rate is given by \cite{Behunin2016,Kamppinen2022}
\begin{equation}\label{eq:dfTTLS2D}
\Delta f_{\rm TTLS,2D} = \frac{\pi}{8 \sqrt{3} } \frac{P_0 \gamma^4}{E^2} \frac{1}{c d^2} \frac{ k_B T }{\hbar^2}.
\end{equation}

As the temperature is increased, the TTLS relaxation rate increases, and becomes approximately equal to the resonance frequency at a threshold temperature $T^*$ \cite{Phillips1987}. Frequency dependence of $T^*$ in different dimensions is discussed in Appendix \ref{AppendixA}.
For temperatures above $T^*$, for all the dimensionalities, TTLS damping saturates to a temperature independent value \cite{Behunin2016}
\begin{equation}\label{eq:dfTTLSHT}
\Delta f_{\rm TTLS,HT} = \frac{\pi f_0}{2} \frac{P_0 \gamma^2}{E}.
\end{equation}
In experiments, saturation of damping sets in at a temperature $T_s$, which is expected to be close to the value $T^*$.

The TTLS contribute to frequency shift of the NEMS devices via relaxation $\delta f_{\rm rel}$ and resonant $\delta f_{\rm res}$ absorption mechanisms, and the total frequency shift is a sum of the two effects $\delta f = \delta f_{\rm rel} + \delta f_{\rm res}$.
The resonant absorption contributes a positive frequency shift \cite{Phillips1987}
\begin{equation} \label{eq:TTLSfshift}
\delta f_{\rm res} = f_0 - f_{0,r} = f_0 \frac{ P_0 \gamma^2 }{ E } \ln\left(\frac{T}{T_r}\right),
\end{equation}
where $f_{0,r}$ is the resonance frequency taken at the reference temperature $T_r$.
In devices where TTLS damping is governed by coupling to bulk phonons, negative frequency shift from relaxation absorption becomes dominant at temperatures $T \gtrsim T^*$, producing a maximum in frequency at approximately $T^*$ \cite{Phillips1987}. 
In 1D systems, contribution from the relaxation absorption to the frequency shift is small, and the frequency is expected to increase past $T^*$ \cite{Seoanez2008,Kamppinen2022}. To our knowledge, prediction for the frequency shift in the 2D case is not found in the literature. 
In our NEMS devices, a maximum in frequency is observed at a temperature $T_m$,
which corresponds to a temperature where negative frequency shift from relaxation absorption of 2D or bulk phonons starts dominating over the positive frequency shift from the resonant absorption mechanism \cite{Kamppinen2022}.

\subsection{Damping from ballistic quasiparticle scattering}

At low temperatures, the mean free path of quasiparticles is long, and they do not interact with each other at length scales smaller than the characteristic size of the device. Thus, their propagation is ballistic.
As a device moves through a superfluid at low temperatures, quasiparticles scatter from its surfaces and exchange momentum with it.
The scattering rate of quasiparticles on either side of the moving device is different, and this results in a net drag force acting on the device \cite{Baym1969,Jager1995,Niemetz2004}.
The difference in scattering rates is proportional to the quasiparticle density $\rho_q$ and the volume swept by the device per unit time. The volume is obtained by integrating the device velocity over the area of the device $$ A_p \dot{x} = \int_{A} \dot{x}(y,z) dy dz .$$ 
Here $\dot{x}(y,z)$ is the velocity at any point on the device surface, and $\dot{x}$ (without explicit position coordinates) refers to the velocity of the paddle.
The average momentum exchanged per quasiparticle is expected to be proportional to the average momentum of the quasiparticles \mbox{$ \langle p_q \rangle = m_q  \langle v_q \rangle $}, where $m_q$ and $\langle v_q \rangle$ are the mass and average speed of the quasiparticles. 
We characterize this proportionality with a scattering efficiency $Q_q$, 
which is the ratio of momentum exchanged with the device per quasiparticle. 
In this notation, the drag force resulting from scattering of ballistic quasiparticles is
\begin{equation}
F_d = Q_q A_p \rho_q \langle v_q \rangle \dot{x},
\end{equation}
and the corresponding damping rate is
\begin{equation}
\Delta f_q = \frac{Q_q}{2 \pi} \frac{A_p}{m} \rho_q \langle v_q \rangle.
\end{equation}
In the so called simple Landau model, the density of phonons is given by \cite{Godfrin2021}
\begin{equation}
 \rho_{p} = \frac{2 \pi^2 k_B^4 T^4}{45 c_p^5 \hbar^3}
\end{equation}
and the same for rotons is
\begin{equation}
 \rho_{r} = \frac{ \hbar k_r^4 (m_r)^{1/2}  }{ 3 \sqrt{2} \pi^{3/2} (k_BT)^{1/2} } \exp \left( - \frac{\Delta_r}{k_B T} \right),
\end{equation}
where \mbox{  $ k_r(\rho_{\rm He}) $ } is the roton wavenumber, \mbox{ $\Delta_r (\rho_{\rm He}) $ } is the roton gap, and
$m_r$ is the effective mass of the roton. 
The phonon and roton densities obtained from the above expressions agree within 10\% with measured densities in Ref. \cite{Godfrin2021} in the regimes where they give the dominant contribution to the damping.
For phonons, \mbox{$\langle v_p \rangle = c_p$}, and the damping rate is
\begin{equation}\label{eq:phonon}
\Delta f_{\rm p} = Q_p \frac{A_p}{m} \frac{\pi (k_B T)^4}{45 \hbar^3 c_p^4}.
\end{equation}
For rotons, $\langle v_r \rangle = \sqrt{2 k_B T / \pi m_r} $, and the damping rate is
\begin{equation}\label{eq:roton}
\Delta f_{\rm r} = Q_r \frac{A_p}{m} \frac{\hbar k_0^4}{6 \pi^3}  \exp{ \left( -\frac{\Delta_r}{k_B T} \right) }.
\end{equation}
Scattering efficiencies close to unity were found for the microspheres studied in Refs. \cite{Jager1995,Niemetz2004}.
Deviation from these values in our devices can arise e.g. due to different shape of the device and different scattering conditions on the device surface (specular or diffuse).

Damping due to $^3$He impurities in the suprefluid $^4$He can be treated in a similar manner as damping from quasiparticles \cite{Rayfield1964}.
The number of $^3$He atoms per unit volume is small, and Maxwell-Boltzmann statistics with the dispersion relation \mbox{$\epsilon_3 = p^2/(2m_3)$} and average velocity \mbox{$v_3 = \sqrt{2 k_B T / \pi m_3}$} is assumed. 
Here $m_3$ is the effective mass of $^3$He particles in superfluid $^4$He, which is approximately 2.4 times the bare atom mass \cite{Yorozu1993}.
The contribution to the damping rate from $^3$He impurities is
\begin{equation} \label{eq:df3}
\Delta f_3 = Q_3 \frac{1}{2 \pi} \frac{A_p}{m}  \sqrt{ \frac{2 k_B T}{\pi m_3} } \rho_3,
\end{equation}
where $\rho_3 $ is the mass density of $^3$He atoms.
Notably, this contribution has the same functional form on temperature as TTLS damping in 1D devices, Eq.~(\ref{eq:dfTTLS1D}).
Assuming $^3$He/$^4$He ratio 1 ppm \cite{Wakita1983}, and $Q_3=1$, we find that the damping predicted by Eq. (\ref{eq:df3}) contributes approximately 2\% increase in the damping at low temperatures, where the major contribution to the damping is given by Eq.~(\ref{eq:dfTTLS1D}).

\begin{figure*}[htb]
\includegraphics[width=0.75\textwidth]{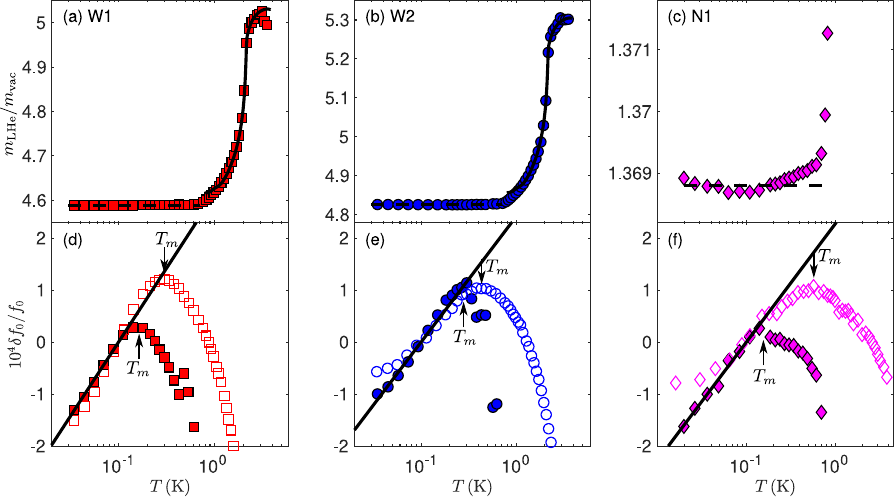}
\caption{ (a--c) Ratio of effective masses in superfluid $^4$He and vacuum, as a function of temperature. The mass ratio is determined from Eq.~(\ref{eq:exp_mass_enhance}) using the measured resonance frequencies in vacuum and in fluid.
Solid lines are fits to Eq.~(\ref{eq:added_mass}). 
Viscous effects are negligible at $T \lesssim \SI{0.7}{\kelvin}$. 
Dashed lines are fits to Eq.~(\ref{eq:added_mass}), neglecting the viscous term.
(d -- f) Detailed view of the resonance frequency shift as a function of temperature, measured in vacuum (empty symbols) and in the fluid (filled symbols). The observed maximum in frequency at temperature $T_m$ is a consequence of combined effect of resonant and relaxation TTLS mechanisms to the frequency shift \cite{Kamppinen2022}. Below $T_m$, the frequency increases logarithmically due to the resonant TTLS mechanism. Lines are fits to Eq.~(\ref{eq:TTLSfshift}). Notably, the slope is practically the same in vacuum and in fluid.
Above $T_m$, negative frequency shift contribution from TTLS relaxation process starts dominating over positive frequency shift contribution from the resonant TTLS mechanism, and frequency decreases. 
In fluid, the resonance frequency is lower,  and relaxation process starts dominating at a lower temperature.}
\label{fig:f0Tdep}
\end{figure*}

\subsection{Viscous flow}

When the mean free path of excitations becomes small compared to the device dimensions, viscous effects become important. The viscous penetration depth, expressed here for the normal fluid component, is \cite{Landau1987}
\begin{equation}
\delta_n = \sqrt{\frac{2 \eta}{\rho_n \omega}},
\end{equation}
where $\eta$ is the dynamic viscosity and $\rho_n$ is the density.
For the device frequency $\sim \SI{30}{\kilo \hertz}$, and at temperatures above the $\lambda$-transition, the viscous penetration depth takes a value $\delta_n \approx \SI{0.5}{\micro \meter}$, which is larger than the device thickness $d$, but smaller than the device width $w$. 
Below the $\lambda$-transition temperature the density of the normal fluid component decreases with decreasing temperature and the viscous penetration depth increases.

We take the width of the beam $w$ as the characteristic size of the body, as suggested in Ref. \cite{Sader1998}. For small oscillation amplitudes ($x_0 \ll d$), damping rate due to the hydrodynamic drag is given by \cite{Blaauwgeers2007}
\begin{equation} \label{eq:viscous_drag}
\Delta f_{h} = \frac{CS}{2m} \sqrt{ \frac{\rho_n \eta f_0}{\pi} },
\end{equation}
where $C$ is a numerical constant of the order of unity, and $S$ is the total surface area of the oscillating body.

\section{Results} \label{section:results}

\begin{figure}
\includegraphics[width=1\columnwidth]{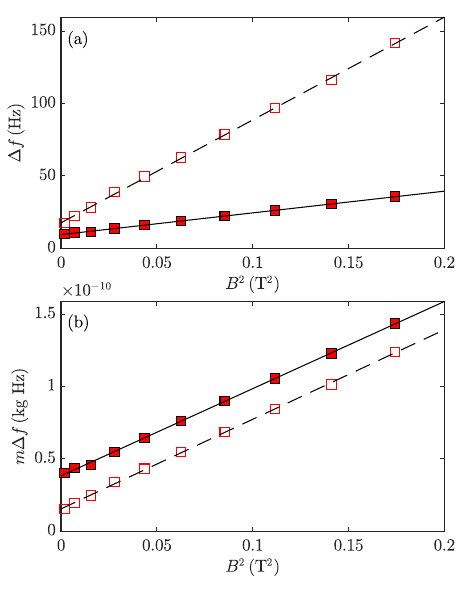}
\caption{(a) Magnetic field dependence of the damping rate of the device W1, measured in vacuum (empty symbols) and in superfluid $^4$He (filled symbols) at $T=\SI{20}{\milli \kelvin}$. Lines are fits to Eq.~(\ref{eq:dfmm}), including in the fitting an additional constant, which is the field-independent damping rate at that particular temperature. The damping rate decreases in the fluid due to the added mass, as expected from Eq. \eqref{eq:massscale}.
(b) Magnetic field dependence of the coefficient $m \Delta f$ which is proportional to the drag force in Eq.~(\ref{eq:drag_force}), in vacuum and fluid.
The slopes of the fit lines are approximately equal, showing that the drag force resulting from magnetomotive damping is independent of frequency, as expected from Eq.~(\ref{eq:mm_estimate}). 
}
\label{fig:mmdep}
\end{figure}

The response of the devices W1, W2, and N1 has been measured as a function of temperature in vacuum and in superfluid $^4$He. 
We start our discussion from the mass enhancement in the fluid and changes in the temperature-independent contributions, as these are used in the subsequent analysis.
The following subsections then describe the temperature-dependent damping contributions from TTLS, ballistic quasiparticles and viscous flow. 
Finally, changes in observed temperature of frequency maxima and temperature of saturation of TTLS damping are discussed.
Physical quantities obtained from the fits in the different temperature regimes are tabulated in Table \ref{table:properties}.

\subsection{Mass enhancement}

The temperature dependence of the resonance frequency, and the ratio of effective masses in vacuum and in superfluid $^4$He for all the three devices W1, W2, and N1 is shown in Fig.~\ref{fig:f0Tdep}. The mass enhancement follows well the theoretical model, Eq.~(\ref{eq:added_mass}), where at the lowest temperatures the last term involving viscosity can be neglected. The geometric constants $\beta$ and $B$ are tabulated in Table~\ref{table:properties}. We were not able to measure device N1 at $T>\SI{0.8}{\kelvin}$ in superfluid $^4$He, as the signal amplitude became too small compared to the electrical background. Reliable fitting of the data for the device N1 in the viscous regime is not possible due to the limited temperature range.

\subsection{Temperature-independent damping}

The magnetomotive damping is determined by measuring the device damping rate $\Delta f$ as a function of the magnetic field, and fitting the measured data to Eq.~(\ref{eq:dfmm}), as shown in Fig. \ref{fig:mmdep}. The prefactors $a_m$ obtained from the fits for all the devices are tabulated in table~\ref{table:properties}.
The fits show that the magnetomotive damping is independent of frequency, as expected from Eq. (\ref{eq:mm_estimate}).
The magnetomotive damping is also independent of temperature \cite{Kamppinen2022}, 
and is subtracted from the measured damping rates for further analysis.

We fit the damping rate with magnetomotive contribution subtracted, $\Delta f - \Delta f_m$, below the saturation temperature, $T<T_s$, to the TTLS model, Eqs.~\eqref{eq:dfTTLS1D} and \eqref{eq:dfTTLS2D}.
For the narrow device N1, no extra contributions compared to Eq.~\eqref{eq:dfTTLS1D} are found. 
For the wide devices W1 and W2, an additional temperature-independent contribution $\Delta f_c$ is included in the fitting. The values $\Delta f_c$ for the devices W1 and W2 are tabulated in Table~\ref{table:properties}.
The damping rates $\Delta f_c$ are almost the same in vacuum and in fluid, but the corresponding drag force 
$F_d = 2 \pi m \Delta f_c \dot{x}$ and dissipated power $P_d = F_d \dot{x}$ are much higher in the fluid, as the device effective mass is enhanced (see Fig. \ref{fig:f0Tdep}).
This is clearly manifested as an offset between the parallel lines in Fig. \ref{fig:mmdep} (b), measured at $T=\SI{20}{\milli \kelvin}$, where temperature-dependent contributions to the damping rate are small.

Acoustic emission to the carrier silicon chip does not seem like a reasonable explanation for the increased dissipated power, as the carrier chip does not support acoustic modes at the frequencies where the wide devices are operated. 
If the increased dissipation was due to acoustic emission to the fluid, we would expect to see it also in the device N1, according to Eq.~(\ref{eq:acoustic_loss}).
In principle, the weak $\Delta f_3 \propto T^{1/2}$ temperature dependence expected from $^3$He impurities could be mistaken for a temperature-independent contribution. 
However, at $T < T_s $ damping rate due to $^3$He impurities $\Delta f_3 \lesssim \SI{0.3}{\hertz}$, obtained from Eq.~\eqref{eq:df3}, is small compared to $\Delta f_c$. 

A possible explanation for the observed temperature-independent contribution is that the entire carrier chip moves due to the elasticity of the glue holding it in place. 
The resonator exerts a periodic force 
\mbox{$F_s = k x$} 
on its support, which drives the motion of the chip.
In fluid, the resonance frequency is lower, and the peak displacement per peak device velocity is higher
($ x_0 / v_0 = (2 \pi f)^{-1} $). Consequently, the force $F_s$ per unit velocity of the device is higher in the fluid.
The oscillations of the chip are expected to be highly damped due to the properties of the glue. 
For highly overdamped resonator, the amplitude of oscillations is expected to increase towards lower frequencies. Thus, both the force $F_s$ driving the dissipative mechanism, and the velocity of the chip and thus the absorbed power, are expected to increase when the device is immersed in the fluid.
We conclude that oscillations of the carrier chip driven by the device motion can qualitatively explain the observed increase in temperature-independent contribution to the damping in fluid.

\subsection{TTLS regime}

\begin{figure}
\includegraphics[width=0.88\columnwidth]{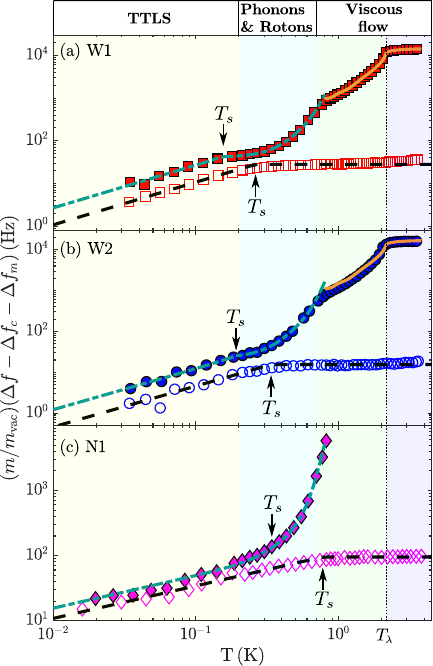}
\caption{Damping rate, $\Delta f$,  of the devices W1 (a), W2 (b), and N1 (c) as a function of temperature, measured in vacuum (empty symbols), and in superfluid $^4$He (filled symbols). The damping rate in the fluid is scaled with the ratio of effective masses $m=m_{\rm LHe}$ and $m_{\rm vac}$. Temperature-independent contributions to the damping have been subtracted. Dashed lines are fits to Eqs.~(\ref{eq:dfTTLS1D} -- \ref{eq:dfTTLSHT}). 
Dash-dot lines are fits to Eqs.~(\ref{eq:dfTTLS1D} -- \ref{eq:dfTTLSHT}), with additional contribution from phonons, Eq.~(\ref{eq:phonon}), and rotons, Eq.~(\ref{eq:roton}), included. 
The saturation temperature $T_s$ marks the transition from temperature-dependent TTLS-damping, governed by Eqs.~(\ref{eq:dfTTLS1D}) and (\ref{eq:dfTTLS2D}), to the saturated, temperature-independent, TTLS-damping regime governed by Eq.~(\ref{eq:dfTTLSHT}).
At $T \gtrsim \SI{0.8}{\kelvin} $ the damping is governed by viscous flow. Full lines are fits to the viscous drag, Eq.~(\ref{eq:viscous_drag}), including the constant TTLS contribution at $T>T_s$. The same model works well in fluid and superfluid phases of $^4$He.}
\label{fig:dfTdep}
\end{figure}

\begin{figure*}
\includegraphics[width = 1.0\textwidth]{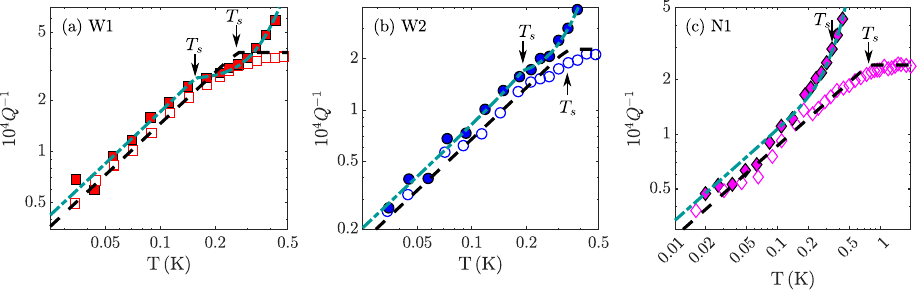}
\caption{Inverse quality factor of the devices W1 (a), W2 (b) and N1 (c) in the TTLS damping regime in vacuum (empty symbols) and fluid (filled symbols).
Temperature-independent contributions to the damping have been subtracted. 
Fit lines are the same as in Fig.~\ref{fig:dfTdep}, but scaled with the frequency of the device.
At $T \lesssim \SI{0.2}{\kelvin}$, contributions from phonons and rotons are negligible, and the quality factors obtained in vacuum and fluid almost collapse on the same line. 
Thus, TTLS damping in the low-temperature regime scales approximately as $\Delta f_{\rm TTLS} \propto f$.
At $T \gtrsim \SI{0.2}{\kelvin}$, contribution from phonons starts visibly affecting the device response in the fluid, and the damping increases beyond the intrinsic values.
} \label{fig:invQ}
\end{figure*}

\begin{table}
	\centering
	\caption{The physical properties of the devices studied in this work. The tabulated resonance frequencies, $f_0$, are taken at $T=\SI{20}{\milli \kelvin}$. 
The parameters $\Delta f_m$ and $\Delta f_c$ are the temperature independent contributions to the damping. 
The values of the TTLS parameters in vacuum, $P_0$ and $\gamma$, are taken from Ref.~\cite{Kamppinen2022}. 
Relative change in phonon-TTLS coupling parameter $\gamma_1$, photon and roton scattering efficiencies $Q_p$ and $Q_r$ and the geometrical constant $C$ describing viscous flow are obtained from the fits shown in Fig. \ref{fig:dfTdep}.
The parameters $B$ and $\beta$ are geometrical parameters describing the mass enhancement resulting from the fluid flow, and are obtained from the fits shown in Fig.~\ref{fig:f0Tdep}.}
\label{table:properties}
\begin{ruledtabular}	
\begin{tabular}{ r | r r r } 
Device 											
& N1 				& W1 				& W2  				\\
$f_{0, \rm vac}$ (\SI{}{\kilo \hertz})   		
& 395.2 				& 73.8 				& 68.0 				\\
$f_{0, \rm LHe}$ (\SI{}{\kilo \hertz})			
& 337.8 				& 34.4 				& 31.0 				\\
$\Delta f_{m,{\rm vac}}$ (\SI{}{\hertz})
& $4 \pm 1$ 			& $124   \pm 2$ 		& $ 67     \pm 2$   	\\
$\Delta f_{m,{\rm LHe}}$ (\SI{}{\hertz})
& $2 \pm 3$ 			& $26.1  \pm 0.5$ 	& $ 12.8   \pm 0.9$  \\
$\Delta f_{c,{\rm vac}} $ (\SI{}{\hertz})
& \hfill - \hfill 	& $12  \pm 3 $		& $4    \pm 3 $ 		\\
$\Delta f_{c,{\rm LHe}} $ (\SI{}{\hertz})
& \hfill - \hfill 	& $8   \pm 1 $	    & $4    \pm 1 $ 		\\
$P_{0{\rm, vac}}$ ($ 10^{-44} \SI{}{ \joule^{-1} \meter^{-3} }$)
& $ 0.49 \pm 0.05 $ 	& $ 7.5 \pm 1.3 $   	& $ 6.3 \pm 1.5 $ 	\\
$\gamma_{\rm vac}$ (\SI{}{\electronvolt})
& $ 2.9 \pm 0.1 $  	& $ 0.93 \pm 0.04 $ 	& $ 0.78 \pm 0.05 $ 	\\
$ \gamma_{1,\rm LHe} / \gamma_{1,\rm vac} $ 
\footnote{ Eqs. \eqref{eq:dfTTLS1D} and \eqref{eq:dfTTLS2D} } 
& $ 1.20 \pm 0.04 $	& $	1.58 \pm 0.28 $	& $ 1.63 \pm 0.39 $  \\
$ \gamma_{1,\rm LHe} / \gamma_{1,\rm vac} $ 
\footnote{Eq. \eqref{eq:dfTTLSHT} }
& $ 1.06 \pm 0.09 $	 & $ 1.78 \pm 0.39 $	& $ 1.82 \pm 0.52 $ \\
$Q_p$	[Eq.~(\ref{eq:phonon})]
& $	2.24 \pm 0.25 $	& $	1.66  \pm 0.08 $	& $	1.57 \pm 0.10 $	\\
$Q_r$	[Eq.~(\ref{eq:roton})]
& $	0.64 \pm 0.08 $	& $	0.11 \pm 0.04 $	& $	0.20 \pm 0.08 $	\\
$C$		[Eq.~(\ref{eq:viscous_drag})]
& \hfill - \hfill	& $ 0.94 \pm 0.01 $ 	&   $ 1.05 \pm 0.03$	\\ 
$B$		[Eq.~(\ref{eq:added_mass})]
& \hfill - \hfill	& $1.61  \pm 0.08 $ 	& 	$ 1.64 \pm 0.08$	\\
$\beta$ 	[Eq.~(\ref{eq:added_mass})]		
& $1.15 \pm 0.06$	& $0.82 \pm 0.04$ 	& 	$ 0.88 \pm 0.04$	\\
\end{tabular}
\end{ruledtabular}
\end{table}

\begin{figure}
\includegraphics[width=0.9\columnwidth]{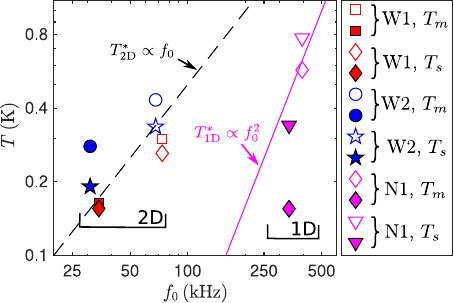}
\caption{Temperature of the frequency maximum, $T_m$, and saturation temperature of damping,   $T_s$, as a function of device frequency, extracted from Figs.~\ref{fig:f0Tdep} and \ref{fig:invQ}, respectively. 
Empty and filled symbols mark data taken in vacuum and fluid, respectively.
The saturation damping temperature $T_s$ is expected to scale similarly as the threshold temperature $T^*$.
The lines are guides to the eye, showing the $T^* \propto f_0$ and $T^* \propto f_0^2$ dependences expected for 2D and 1D devices, respectively (see Appendix \ref{AppendixA}). 
The change in $T_s$ as a function of frequency for the 1D device N1 is much steeper than for the 2D devices W1 and W2, as expected from the theory.
The temperature of the frequency maximum, $T_m$, is expected to scale similarly as $T^*$ for bulk amorphous insulators. For the 1D and 2D devices, it is expected to mark temperature where either 2D or 3D phonon processes start dominating the TTLS relaxation process.
We observe that the value for the frequency maximum $T_m$ scales similarly as $T_s$ for the wide devices $W1$ and $W2$, but much more steeply for the narrow device $N1$.}
\label{fig:Tstar}
\end{figure}

The damping rate as a function of temperature in vacuum and in fluid for the three devices is shown in Fig.~\eqref{fig:dfTdep}. 
The temperature-independent contributions $\Delta f_m$ and $\Delta f_c$ discussed in the previous section, and tabulated in Table~\ref{table:properties} have been reduced from the data, and the damping rate in the fluid has been scaled with the effective mass $m/m_{\rm vac}$ shown in Fig. \ref{fig:f0Tdep}. 

Below \SI{0.2}{\kelvin}, contribution from ballistic quasiparticles to the damping rate is negligible, and damping is governed by the intrinsic TTLS damping mechanism. 
As seen in Fig.~\ref{fig:dfTdep}, the damping rate has the same functional form in vacuum and in fluid in this temperature regime, i.e.  
$\Delta f \propto T$ for the wide 2D-devices W1 and W2, and $\Delta f \propto T^{1/2}$ for the narrow 1D-device N1. 
However, the drag force and the corresponding dissipated power are higher in the fluid, manifested as an offset between the parallel lines in the TTLS regime. This corresponds to an increase in the product $P_0 \gamma^4$ in Eqs. \eqref{eq:dfTTLS1D} and \eqref{eq:dfTTLS2D}.
Transition to the saturated damping regime is somewhat masked by the contribution from phonons and rotons in the fluid, but is still visible as a small kink in the data around $T_s$.
Also the saturated damping regime shows increased dissipation, indicating an increase in the product $P_0 \gamma^2$ in Eq.~\eqref{eq:dfTTLSHT}.

We note that the value $P_0$ describing the TTLS density of states is a property of the TTLS distribution in the material of the device and is not expected to change with immersion of the device into fluid. It is also expected to be energy-independent in the relevant range.
While a possible energy dependence $ P_0 \propto \epsilon^\mu $ in some mesoscopic systems has been suggested,
it should manifest as deviations from the observed temperature dependences of damping, producing  $ \Delta f_{\rm TTLS, 1D} \propto T^{1/2 + \mu} $, $ \Delta f_{\rm TTLS, 2D} \propto T^{1+\mu} $ and $ \Delta f_{\rm TTLS,HT} \propto T^\mu $ \cite{Behunin2016}. 
Our experimental data are best fit with $ \mu = 0.0 \pm 0.1 $. 
Thus, we assume that $P_0$ keeps its vacuum value in the liquid. 
Another parameter in Eqs.~\eqref{eq:dfTTLS1D} and \eqref{eq:dfTTLS2D} which potentially changes when the devices are immersed in the liquid is the sound velocity $c$.
We have considered the effect of added mass from the fluid on the phonon dispersion relation in Appendix~\ref{AppendixB}, but these corrections have proven to be small.

As a result, we are left with TTLS-phonon coupling $\gamma$ as responsible for the change of the damping in helium. Before further analysis, we note that TTLS interact with phonon modes at very different frequencies: First, the low-frequency mode corresponding to the device oscillations, and, in the case of relaxation absorption, with phonon bath at high frequencies (see Appendix \ref{AppendixB}). 
In derivation of Eqs.~(\ref{eq:dfTTLS1D}--\ref{eq:TTLSfshift}) it is assumed that the coupling $\gamma$ is the same for all modes. We generalize the approach and allow different couplings: $\gamma_1$ for the device-frequency mode and $\gamma_2$ for high-frequency modes. Then in Eqs.~(\ref{eq:dfTTLS1D}--\ref{eq:dfTTLS2D}) $\gamma^4$ is replaced by $\gamma_1^2 \gamma_2^2$, and in Eq.~\eqref{eq:dfTTLSHT} $\gamma^2$ is replaced by $\gamma_1^2$. In vacuum, $ \gamma_1 = \gamma_2 = \gamma_{\rm vac} $. 
In liquid, we expect $\gamma_2$ not to change significantly (see Appendix \ref{AppendixB} and below), and we determine value of $\gamma_1$ from the low-temperature behavior of damping,
Eqs.~(\ref{eq:dfTTLS1D} -- \ref{eq:dfTTLS2D}), or from the saturated value of damping, Eq.~\eqref{eq:dfTTLSHT}, assuming $P_0$ and $\gamma_2$ to keep their values obtained in vacuum.
Results are shown in Table~\ref{table:properties}, and both methods give approximately equal values of $\gamma_1$ in helium.
Remarkably, we empirically find scaling $ \gamma_1 \propto f_0^{-1/2} $. Such scaling implies that the inverse quality factor $Q^{-1}$ does not change on immersion of the device to the ideal fluid. This property is verified in Fig.~\ref{fig:invQ}.

We note that the frequency shift data shown in Fig.~\ref{fig:f0Tdep} suggest that the 
product $P_0 \gamma^2$ does not change when the devices are immersed in the fluid. 
Here, the slope of the frequency shift at temperatures $T<T_m$ is proportional to $P_0 \gamma^2$ according to Eq.~\eqref{eq:TTLSfshift}. 
TTLS at energies close to $k_B T$ give the dominant contribution to the frequency shift \cite{Phillips1987}. 
Thus, the $\gamma$ obtained from the frequency shift is relevant for phonons at high thermal frequencies, that is $\gamma_2$ in our notation. 
The frequency data thus supports our assumption that $\gamma_2$ does not change when the devices are immersed to fluid.

The temperatures $T_s$ (temperature of onset of TTLS damping saturation in Fig.~\ref{fig:dfTdep}) and $T_m$ (temperature of the frequency maxima in Fig.~\ref{fig:f0Tdep}), are plotted as a function of the device frequency in Fig.~\ref{fig:Tstar}. 
The values of $T_s$ are expected to scale as $T^*$ for the particular device, when the frequency of the device changes. Also the values $T_m$ are expected to decrease with frequency, but here the distinction between different regimes is not so clear (see Section \ref{section:TTLS_theory}).
For the 1D-device N1, $T_s$ and $T_m$ change more rapidly as a function of frequency than for the 2D-devices W1 and W2, as expected for the change in $T^*$ (see Appendix \ref{AppendixA}).

\subsection{Ballistic regime}

The roton and phonon scattering efficiencies $Q_p$ and $Q_r$ for the different devices are extracted from the fits shown in Fig.~\ref{fig:dfTdep}, and the values are tabulated in Table~\ref{table:properties}. The phonon scattering efficiencies $Q_p \sim 2$ for all the devices. For spheres oscillating in superfluid $^4$He, this value has been found to be close to unity \cite{Niemetz2004}, where the difference is perhaps explained by the different geometry of the device, or different scattering conditions (specular or diffuse) on the device surface. For rotons, we find that the scattering efficiency $Q_r$ is smaller than unity for all the devices. Due to the differences in $Q_p$ and $Q_r$ between the devices, calibration is required if one wishes to use the devices for thermometry in superfluid $^4$He in the ballistic regime.

\subsection{Viscous regime}

In the viscous regime, above \SI{800}{\milli \kelvin}, where data on fluid properties is readily available, we  fit the device W1 and W2 response to Eqs.~(\ref{eq:added_mass}) and (\ref{eq:viscous_drag}), as shown in Figs.~\ref{fig:f0Tdep} and \ref{fig:dfTdep}. The parameters $B$, $\beta$, and $C$ are tabulated in Table~\ref{table:properties}. 
The parameter $C$ for both the devices is very close to unity, and perhaps these devices could be used for thermo- or viscometry in a fluid even without free parameters. 
It is also notable that the variation of the parameters $B$ and $\beta$ between the devices are small, so the frequency could be used as an alternative measuring technique, perhaps without free parameters.

\section{Conclusions}

We have measured the damping and frequency shift of three NEMS resonators in vacuum and in superfluid $^4$He at temperatures from \SI{20}{\milli \kelvin} to \SI{4}{\kelvin}. Our measurements span over four orders of magnitude in damping, enabling rigorous test on the existing models describing device-intrinsic damping and device-fluid interactions. 

The dominant device-intrinsic damping mechanism in our devices is TTLS damping. 
Beyond nanoelectromechanical resonators, TTLS affect noise, dissipation, and decoherence in a wide range of quantum-limited measurements, e.g. in qubits and optomechanical systems. 
Immersing mechanical resonators in superfluid $^4$He gave us the possibility to study TTLS damping in a  setting, where the frequency of the mechanical mode could be reduced in situ by up to about 50\% via mass enhancement from the fluid without adding extra dissipation.
We find that the damping rate due to TTLS scales approximately as $\Delta f \propto f_0$, while scaling $\Delta f \propto 1/m \propto f_0^2$ is expected from mass loading only. Thus, intrinsic drag due to TTLS is increased in the fluid. 
We attribute the increased damping to a TTLS-phonon coupling parameter $\gamma_1$ taken at the device frequency $f_0$, and scaling approximately as $\gamma_1 \propto f_0 ^{-1/2}$.
In future, systematic studies on the parameter $\gamma_1$ as a function of device frequency could be done by measurements at various fluid densities, which allows further tuning of the frequency of the devices by up to 7\%.

Another important damping mechanism in our devices is magnetomotive damping. We find that magnetomotive damping is independent of frequency in the frequency range \SIrange{30}{400}{\kilo \hertz}, and the corresponding damping rate is inversely proportional to the effective mass of the resonator. In addition, changing the frequency allowed us to study the previously unidentified temperature-independent damping mechanism in our devices \cite{Kamppinen2022}, and we propose overdamped oscillations of the carrier chip as a possible explanation for the observed damping. 

In our devices, large frequency tuning by mass loading from fluid is achieved by making devices with large aspect ratio, where the ratio of beam width to thickness is $\sim 100$ in wider devices.
As mechanical resonators are the most sensitive to forces acting on the device at the mechanical resonance frequency, the frequency tuning has many potential applications in studying effects that occur at specific frequencies. Examples of such effects are 
resonant Kelvin waves on quantized vortices in superfluids \cite{Vinen2003,Eltsov2020}, acoustic modes in cavities \cite{Shkarin2019}, and vortex-core-bound states in superfluid $^3$He \cite{Kopnin1991}. Beyond superfluids, the frequency tuning could be utilized for instance in NEMS based nuclear magnetic resonance measurements \cite{Myller2014}.

At $T>\SI{0.2}{\kelvin}$ contributions from thermal excitations of $^4$He, namely phonons and rotons, increase the damping of the NEMS devices. Good agreement with existing theory is found, but with some differences in the scattering efficiencies found between the devices, perhaps due to different surface roughness conditions. The devices are very sensitive to the quasiparticles due to the large surface area and small mass, and the calibrated devices could be used for precise thermometry in superfluid $^4$He. As the temperature is increased furter, above $T>\SI{0.8}{\kelvin}$, viscous effects become important. The obtained geometrical parameters are close to unity, as expected, and agreement between different devices is good. The same geometrical parameters work in the normal and superfluid states of $^4$He.
This shows that similar devices could be used for viscometry, and thermometry in superfluid $^4$He, with good precision without free parameters. 

\begin{acknowledgments}

We thank Henri Godfrin for useful comments.
We acknowledge the technical support from Micronova Nanofabrication Centre of VTT.
This work has been supported by the European Research Council (ERC) under the
European Union’s Horizon 2020 research and innovation programme (Grant Agreement No. 694248) and by Academy of Finland (Grant No. 332964). The research leading to these results has received funding from the European Union's Horizon 2020 research and innovation programme under Grant Agreement No. 824109. The experiments were performed at the Low Temperature Laboratory, which is a part of the OtaNano research infrastructure of Aalto University and of the EU H2020 European Microkelvin Platform.
T. K. acknowledges financial support from the Finnish Cultural Foundation (Grants No. 00190453, No. 00201211, and No. 00212577).

\end{acknowledgments}

\appendix

\section{TTLS saturation temperature $T^*$ in different dimensions} \label{AppendixA}

The spatially and orientationally averaged TTLS relaxation rate is \cite{Behunin2016,Kamppinen2022}
\begin{equation} \label{eq:T1}
\langle \tau_1^{-1} (\epsilon) \rangle_V  = \frac{1}{V} g( \epsilon ) \frac{ \Delta_0^2 }{ \epsilon } \frac{ \pi \gamma^2 }{ E \hbar^2 } \coth \left( \frac{\epsilon}{ 2 k_B T } \right),
\end{equation}
where $g( \epsilon )$ is the phonon density of states, and $\epsilon = \sqrt{\Delta_0^2 + \Delta^2} $ is the TTLS energy, $\Delta_0$ is the tunneling strength and $\Delta$ is the double well asymmetry.
The density of states for flexural phonons in 1D and 2D are given by \mbox{$g_{\rm 1D}( \epsilon ) \propto \epsilon^{-1/2}$} and \mbox{$g_{\rm 2D}( \epsilon ) \propto \epsilon^0$}, respectively \cite{Behunin2016,Kamppinen2022}, and for bulk phonons \mbox{$g_{\rm 3D}( \epsilon ) = \epsilon^2$} \cite{Phillips1987}. 
The minimum relaxation time $\tau_{1,{\rm min}}$ is obtained for the TTLS states with $\epsilon = \Delta_0$. Inserting the density of states in Eq.~(\ref{eq:T1}), we get
\begin{equation} \label{eq:T1dims}
\begin{split}
\tau_{1,{\rm min, 1D}} & \propto \epsilon^{-1/2} \coth^{-1} \left( \frac{\epsilon}{ 2 k_B T } \right) \propto \epsilon^{1/2}|_{\epsilon \ll k_B T} \\
\tau_{1,{\rm min, 2D}} & \propto \epsilon^{-1} \coth^{-1} \left( \frac{\epsilon}{ 2 k_B T } \right) \propto \epsilon^0|_{\epsilon \ll k_B T} \\
\tau_{1,{\rm min, 3D} } & \propto \epsilon^{-3} \coth^{-1} \left( \frac{\epsilon}{ 2 k_B T } \right) \propto \epsilon^{-2}|_{\epsilon \ll k_B T},
\end{split}
\end{equation}
where only the states up to $\epsilon \approx k_B T$ are relevant \cite{Kamppinen2022}. For the states with $\epsilon = k_B T$, we have
\begin{equation}
\begin{split}
\tau_{1,{\rm min, 1D} } (\epsilon = k_B T) & \propto (k_B T)^{-1/2}  \\
\tau_{1,{\rm min, 2D} } (\epsilon = k_B T) & \propto (k_B T)^{-1}  \\
\tau_{1,{\rm min 3D} }  (\epsilon = k_B T) & \propto (k_B T)^{-3} .
\end{split}
\end{equation}
We require that $$2 \pi f_0 \tau_{1,{\rm min}}(\epsilon = k_B T^*) = 1,$$ which results in
\begin{equation}
\begin{split}
T^*_{\rm 1D} & \propto f_0^2 \\
T^*_{\rm 2D} & \propto f_0 \\
T^*_{\rm 3D} & \propto f_0^{1/3}.
\end{split}
\end{equation}
The expressions are useful when comparing changes in $T^*$ within devices of the same dimensionality, but some caution is advised if the same are to be applied to devices of different dimensionalities. 
It follows from the right-hand-side terms in Eq.~(\ref{eq:T1dims}) that in 1D, 
\mbox{$\tau_{1,{\rm min,1D}}(\epsilon = k_B T^*_{\rm 1D})$} is a maximum, i.e. the TTLS with $\epsilon < k_B T$ have shorter relaxation times. Thus, $T^*_{\rm 1D}$ marks the temperature above which practically all TTLS have $ 2 \pi f_0 \tau_1 \lesssim 1 $. 
Similarly, it follows that in 2D \mbox{$\tau_{1,{\rm min,2D}}(\epsilon)$} is almost independent of the energy and $T^*_{\rm 2D}$ marks the temperature where practically all TTLS have $ 2 \pi f_0 \tau_1 \sim 1$. In 3D,  \mbox{$\tau_{1,{\rm min,3D}}(\epsilon = k_B T^*_{\rm 3D})$} is a minimum, i.e. the TTLS with $\epsilon < k_B T$ have longer relaxation times and marks the temperature below which practically all TTLS have $ 2 \pi f_0 \tau_1 \gtrsim 1 $.

\section{Added mass contribution to phonon dispersion relation} \label{AppendixB}

Analytical expressions for TTLS damping in reduced dimensions usually rely on expressions derived from phonon dispersion relation for a simple geometry, such as a beam or plate in vacuum \cite{Seoanez2008,Behunin2016,Kamppinen2022}. 
Here, we extend these models from the simplest case of a beam in vacuum to a beam in fluid, taking in to account change in the phonon dispersion relation due to increase in the effective mass. 

The dispersion relation for flexural phonons in a rectangular cantilever beam is given by
\begin{equation} \label{eq:phonondispersion}
\omega = k_{\rm ph}^2 \sqrt{\frac{E I_x}{\rho_{\rm Al} w d}},
\end{equation}
where $k_{\rm ph}$ is the phonon wavenumber, $E$ is the Young's modulus, $I_x = w d^3/12$ is the second moment of inertia, $w$ is the beam width, and $d$ is the thickness.
Strictly speaking, Eq.~\eqref{eq:phonondispersion} is valid only in vacuum, and in fluid it should be modified by the presence of the fluid via mass loading. 

The flexural phonon frequencies given by Eq. (\ref{eq:phonondispersion}) are closely related to the natural frequencies of a fixed-free  cantilever beam.
For a beam of length $H$, width $w$, and thickness $d$, the natural frequencies are given by \cite{LandauLifshitz1986}
\begin{equation} \label{eq:eigenmodes}
 \omega_0 = \frac{k_n}{H^2} \sqrt{\frac{E I_x}{\rho_{\rm Al} w d}},
\end{equation}
where $k_n$ are the roots of the equation $\cos( \sqrt{k_n} ) \cosh( \sqrt{k_n}) + 1 = 0$, where $n$ is the mode number. For example, for the first three modes $k_1 \approx 3.52$, $k_2 \approx 22.0 $, and $k_3 \approx 61.7$. The free end of the beam is an anti-node, and consequently the eigenmodes are odd multiples of the quarter wavelength of the corresponding flexural phonons. 
This is seen by setting 
$ k_{\rm ph} = (2n-1)\pi/(2H) $ in Eq.~(\ref{eq:phonondispersion}), which produces 
values close to that of Eq.~(\ref{eq:eigenmodes}), 
with decreasing deviation as the mode number $n$ increases.

When a device is immersed in a fluid, its effective mass increases, and its resonance frequency decreases according to Eq.~\eqref{eq:f0}.
Our experiments show that the mass-enhancement at low temperatures results solely from the potential flow of the fluid (second term in Eq. (\ref{eq:added_mass}), on the right-hand side). 
The parameter $\beta$ describing the potential flow is by first principles obtained by integrating the fluid velocity field around the device \cite{Landau1987}.
Due to the close resemblance between flexural phonon modes and the mechanical eigenmodes of the device, we believe that flexural phonons with sufficiently low frequencies should scale similarly as the mechanical mode, when immersed in the fluid  
\begin{equation}
 \omega_{\rm LHe} = \left( \frac{ \omega_{0, {\rm LHe}} }{ \omega_{0, {\rm vac}} } \right) \omega.
\end{equation}
In terms of the parameters appearing in Eqs.~\eqref{eq:dfTTLS1D}, \eqref{eq:dfTTLS2D}, \eqref{eq:phonondispersion} and \eqref{eq:eigenmodes}, 
the change in the frequency can be conveniently incorporated in an effective speed of sound $ c_{\rm LHe} = (\omega_{0,{\rm LHe}}/\omega_{0,{\rm vac}}) c $.

At sufficiently high frequencies, the velocity along the beam varies at a length scale which is shorter than the beam width, which sets the relevant hydrodynamic length scale \cite{Sader1998}. In this case, fluid can take a shortcut by moving from antinode to antinode, rather than around the beam. 
With increasing frequency the distance between antinodes decreases, and we expect that the mass enhancement from the potential flow diminishes.
For the wide devices W1 and W2, the flexural phonon wavelength becomes smaller than the beam width at around \SI{5}{\mega \hertz}, and for the narrow device N1, at around \SI{1}{\giga \hertz}. 
At higher phonon frequencies, the vacuum phonon dispersion relation given by Eq. (\ref{eq:phonondispersion}) is expected to hold.

Eqs. (\ref{eq:dfTTLS1D}) and (\ref{eq:dfTTLS2D}) describing TTLS losses in 1D and 2D devices, respectively, are derived from an integral of the form \cite{Behunin2016}
\begin{equation}
 \Delta f \propto \frac{1}{k_B T} \int_0^{ \infty } d \epsilon \left[ \epsilon g(\epsilon) {\rm csch} \left( \frac{\epsilon}{k_B T} \right) \right],
\end{equation}
where $ \epsilon $ is the TTLS energy, and $g(\epsilon)$ is the phonon density of states introduced in Appendix \ref{AppendixA}.
The ${\rm csch}(\epsilon/ k_B T )$-term imposes a temperature-dependent cutoff frequency.
For the wide 2D devices W1 and W2, the dominant contribution to this integral at $T>\SI{1}{\milli \kelvin}$ comes from phonons with frequency above \SI{5}{\mega \hertz}. Thus, we expect that mass enhancement in fluid is irrelevant for the expression given in Eq. (\ref{eq:dfTTLS2D}).
For the narrow 1D device N1, a substantial fraction to the integral is contributed by phonon states below \SI{1}{\giga \hertz}, and they give the dominant contribution to the integral at $T<\SI{0.2}{\kelvin}$. 
The relative importance of the mass-scaled phonon frequencies decreases with increasing temperature due to the temperature-dependent cutoff frequency.
The maximum relative error in damping resulting from using the vacuum speed of sound in Eq. (\ref{eq:dfTTLS1D}) for the device N1 is
$$ 1 - \sqrt{  c /  c_{\rm LHe} } = 1 - \sqrt{ f_{0,{\rm vac}} / f_{0,{\rm LHe} } } \approx -8\%. $$ 
The expected effect for the narrow device N1 is on par with other error sources.

\bibliography{NEMS4HeTdep}  

\begin{thebibliography}{38}%
\makeatletter
\providecommand \@ifxundefined [1]{%
 \@ifx{#1\undefined}
}%
\providecommand \@ifnum [1]{%
 \ifnum #1\expandafter \@firstoftwo
 \else \expandafter \@secondoftwo
 \fi
}%
\providecommand \@ifx [1]{%
 \ifx #1\expandafter \@firstoftwo
 \else \expandafter \@secondoftwo
 \fi
}%
\providecommand \natexlab [1]{#1}%
\providecommand \enquote  [1]{``#1''}%
\providecommand \bibnamefont  [1]{#1}%
\providecommand \bibfnamefont [1]{#1}%
\providecommand \citenamefont [1]{#1}%
\providecommand \href@noop [0]{\@secondoftwo}%
\providecommand \href [0]{\begingroup \@sanitize@url \@href}%
\providecommand \@href[1]{\@@startlink{#1}\@@href}%
\providecommand \@@href[1]{\endgroup#1\@@endlink}%
\providecommand \@sanitize@url [0]{\catcode `\\12\catcode `\$12\catcode
  `\&12\catcode `\#12\catcode `\^12\catcode `\_12\catcode `\%12\relax}%
\providecommand \@@startlink[1]{}%
\providecommand \@@endlink[0]{}%
\providecommand \url  [0]{\begingroup\@sanitize@url \@url }%
\providecommand \@url [1]{\endgroup\@href {#1}{\urlprefix }}%
\providecommand \urlprefix  [0]{URL }%
\providecommand \Eprint [0]{\href }%
\providecommand \doibase [0]{https://doi.org/}%
\providecommand \selectlanguage [0]{\@gobble}%
\providecommand \bibinfo  [0]{\@secondoftwo}%
\providecommand \bibfield  [0]{\@secondoftwo}%
\providecommand \translation [1]{[#1]}%
\providecommand \BibitemOpen [0]{}%
\providecommand \bibitemStop [0]{}%
\providecommand \bibitemNoStop [0]{.\EOS\space}%
\providecommand \EOS [0]{\spacefactor3000\relax}%
\providecommand \BibitemShut  [1]{\csname bibitem#1\endcsname}%
\let\auto@bib@innerbib\@empty
\bibitem [{\citenamefont {Li}(2007)}]{LiMo2007}%
  \BibitemOpen
  \bibfield  {author} {\bibinfo {author} {\bibfnamefont {M.}~\bibnamefont
  {Li}},\ }\bibfield  {title} {\bibinfo {title} {Ultra-sensitive {NEMS}-based
  cantilevers for sensing, scanned probe and very high-frequency
  applications},\ }\href {https://doi.org/10.1038/nnano.2006.208} {\bibfield
  {journal} {\bibinfo  {journal} {Nature Nanotechnology}\ }\textbf {\bibinfo
  {volume} {2}},\ \bibinfo {pages} {114} (\bibinfo {year} {2007})}\BibitemShut
  {NoStop}%
\bibitem [{\citenamefont {Müller}(2014)}]{Myller2014}%
  \BibitemOpen
  \bibfield  {author} {\bibinfo {author} {\bibfnamefont {C.}~\bibnamefont
  {Müller}},\ }\bibfield  {title} {\bibinfo {title} {Nuclear magnetic
  resonance spectroscopy with single spin sensitivity},\ }\href
  {https://doi.org/10.1038/ncomms5703} {\bibfield  {journal} {\bibinfo
  {journal} {Nature Communications}\ }\textbf {\bibinfo {volume} {5}},\
  \bibinfo {pages} {4703} (\bibinfo {year} {2014})}\BibitemShut {NoStop}%
\bibitem [{\citenamefont {Guthrie}(2021)}]{Guthrie2021}%
  \BibitemOpen
  \bibfield  {author} {\bibinfo {author} {\bibfnamefont {A.}~\bibnamefont
  {Guthrie}},\ }\bibfield  {title} {\bibinfo {title} {Nanoscale real-time
  detection of quantum vortices at millikelvin temperatures},\ }\href
  {https://doi.org/10.1038/s41467-021-22909-3} {\bibfield  {journal} {\bibinfo
  {journal} {Nature Communications}\ }\textbf {\bibinfo {volume} {12}},\
  \bibinfo {pages} {2645} (\bibinfo {year} {2021})}\BibitemShut {NoStop}%
\bibitem [{\citenamefont {Kopnin}\ and\ \citenamefont
  {Salomaa}(1991)}]{Kopnin1991}%
  \BibitemOpen
  \bibfield  {author} {\bibinfo {author} {\bibfnamefont {N.~B.}\ \bibnamefont
  {Kopnin}}\ and\ \bibinfo {author} {\bibfnamefont {M.~M.}\ \bibnamefont
  {Salomaa}},\ }\bibfield  {title} {\bibinfo {title} {Mutual friction in
  superfluid $^{3}\mathrm{He}$: Effects of bound states in the vortex core},\
  }\href {https://doi.org/10.1103/PhysRevB.44.9667} {\bibfield  {journal}
  {\bibinfo  {journal} {Phys. Rev. B}\ }\textbf {\bibinfo {volume} {44}},\
  \bibinfo {pages} {9667} (\bibinfo {year} {1991})}\BibitemShut {NoStop}%
\bibitem [{\citenamefont {Kamppinen}\ \emph {et~al.}(2022)\citenamefont
  {Kamppinen}, \citenamefont {M\"akinen},\ and\ \citenamefont
  {Eltsov}}]{Kamppinen2022}%
  \BibitemOpen
  \bibfield  {author} {\bibinfo {author} {\bibfnamefont {T.}~\bibnamefont
  {Kamppinen}}, \bibinfo {author} {\bibfnamefont {J.~T.}\ \bibnamefont
  {M\"akinen}},\ and\ \bibinfo {author} {\bibfnamefont {V.~B.}\ \bibnamefont
  {Eltsov}},\ }\bibfield  {title} {\bibinfo {title} {Dimensional control of
  tunneling two-level systems in nanoelectromechanical resonators},\ }\href
  {https://doi.org/10.1103/PhysRevB.105.035409} {\bibfield  {journal} {\bibinfo
   {journal} {Phys. Rev. B}\ }\textbf {\bibinfo {volume} {105}},\ \bibinfo
  {pages} {035409} (\bibinfo {year} {2022})}\BibitemShut {NoStop}%
\bibitem [{\citenamefont {Behunin}\ \emph {et~al.}(2016)\citenamefont
  {Behunin}, \citenamefont {Intravaia},\ and\ \citenamefont
  {Rakich}}]{Behunin2016}%
  \BibitemOpen
  \bibfield  {author} {\bibinfo {author} {\bibfnamefont {R.~O.}\ \bibnamefont
  {Behunin}}, \bibinfo {author} {\bibfnamefont {F.}~\bibnamefont {Intravaia}},\
  and\ \bibinfo {author} {\bibfnamefont {P.~T.}\ \bibnamefont {Rakich}},\
  }\bibfield  {title} {\bibinfo {title} {Dimensional transformation of
  defect-induced noise, dissipation, and nonlinearity},\ }\href
  {https://doi.org/10.1103/PhysRevB.93.224110} {\bibfield  {journal} {\bibinfo
  {journal} {Physical Review B}\ }\textbf {\bibinfo {volume} {93}},\ \bibinfo
  {pages} {224110} (\bibinfo {year} {2016})}\BibitemShut {NoStop}%
\bibitem [{\citenamefont {Ithier}\ \emph {et~al.}(2005)\citenamefont {Ithier},
  \citenamefont {Collin}, \citenamefont {Joyez}, \citenamefont {Meeson},
  \citenamefont {Vion}, \citenamefont {Esteve}, \citenamefont {Chiarello},
  \citenamefont {Shnirman}, \citenamefont {Makhlin}, \citenamefont {Schriefl},\
  and\ \citenamefont {Sch\"on}}]{Ithier2005}%
  \BibitemOpen
  \bibfield  {author} {\bibinfo {author} {\bibfnamefont {G.}~\bibnamefont
  {Ithier}}, \bibinfo {author} {\bibfnamefont {E.}~\bibnamefont {Collin}},
  \bibinfo {author} {\bibfnamefont {P.}~\bibnamefont {Joyez}}, \bibinfo
  {author} {\bibfnamefont {P.~J.}\ \bibnamefont {Meeson}}, \bibinfo {author}
  {\bibfnamefont {D.}~\bibnamefont {Vion}}, \bibinfo {author} {\bibfnamefont
  {D.}~\bibnamefont {Esteve}}, \bibinfo {author} {\bibfnamefont
  {F.}~\bibnamefont {Chiarello}}, \bibinfo {author} {\bibfnamefont
  {A.}~\bibnamefont {Shnirman}}, \bibinfo {author} {\bibfnamefont
  {Y.}~\bibnamefont {Makhlin}}, \bibinfo {author} {\bibfnamefont
  {J.}~\bibnamefont {Schriefl}},\ and\ \bibinfo {author} {\bibfnamefont
  {G.}~\bibnamefont {Sch\"on}},\ }\bibfield  {title} {\bibinfo {title}
  {Decoherence in a superconducting quantum bit circuit},\ }\href
  {https://doi.org/10.1103/PhysRevB.72.134519} {\bibfield  {journal} {\bibinfo
  {journal} {Phys. Rev. B}\ }\textbf {\bibinfo {volume} {72}},\ \bibinfo
  {pages} {134519} (\bibinfo {year} {2005})}\BibitemShut {NoStop}%
\bibitem [{\citenamefont {de~Lépinay}\ \emph {et~al.}(2021)\citenamefont
  {de~Lépinay}, \citenamefont {Ockeloen-Korppi}, \citenamefont {Woolley},\
  and\ \citenamefont {Sillanpää}}]{Laure2021}%
  \BibitemOpen
  \bibfield  {author} {\bibinfo {author} {\bibfnamefont {L.~M.}\ \bibnamefont
  {de~Lépinay}}, \bibinfo {author} {\bibfnamefont {C.~F.}\ \bibnamefont
  {Ockeloen-Korppi}}, \bibinfo {author} {\bibfnamefont {M.~J.}\ \bibnamefont
  {Woolley}},\ and\ \bibinfo {author} {\bibfnamefont {M.~A.}\ \bibnamefont
  {Sillanpää}},\ }\bibfield  {title} {\bibinfo {title} {Quantum
  mechanics-free subsystem with mechanical oscillators},\ }\href
  {https://doi.org/10.1126/science.abf5389} {\bibfield  {journal} {\bibinfo
  {journal} {Science}\ }\textbf {\bibinfo {volume} {372}},\ \bibinfo {pages}
  {625} (\bibinfo {year} {2021})}\BibitemShut {NoStop}%
\bibitem [{\citenamefont {Seo\'anez}\ \emph {et~al.}(2008)\citenamefont
  {Seo\'anez}, \citenamefont {Guinea},\ and\ \citenamefont
  {Castro~Neto}}]{Seoanez2008}%
  \BibitemOpen
  \bibfield  {author} {\bibinfo {author} {\bibfnamefont {C.}~\bibnamefont
  {Seo\'anez}}, \bibinfo {author} {\bibfnamefont {F.}~\bibnamefont {Guinea}},\
  and\ \bibinfo {author} {\bibfnamefont {A.~H.}\ \bibnamefont {Castro~Neto}},\
  }\bibfield  {title} {\bibinfo {title} {Surface dissipation in
  nanoelectromechanical systems: Unified description with the standard
  tunneling model and effects of metallic electrodes},\ }\href
  {https://doi.org/10.1103/PhysRevB.77.125107} {\bibfield  {journal} {\bibinfo
  {journal} {Phys. Rev. B}\ }\textbf {\bibinfo {volume} {77}},\ \bibinfo
  {pages} {125107} (\bibinfo {year} {2008})}\BibitemShut {NoStop}%
\bibitem [{\citenamefont {Blaauwgeers}\ \emph {et~al.}(2007)\citenamefont
  {Blaauwgeers}, \citenamefont {Blazkova}, \citenamefont {Človečko},
  \citenamefont {Eltsov}, \citenamefont {de~Graaf}, \citenamefont {Hosio},
  \citenamefont {Krusius}, \citenamefont {Schmoranzer}, \citenamefont
  {Schoepe}, \citenamefont {Skrbek}, \citenamefont {Skyba}, \citenamefont
  {Solntsev},\ and\ \citenamefont {Zmeev}}]{Blaauwgeers2007}%
  \BibitemOpen
  \bibfield  {author} {\bibinfo {author} {\bibfnamefont {R.}~\bibnamefont
  {Blaauwgeers}}, \bibinfo {author} {\bibfnamefont {M.}~\bibnamefont
  {Blazkova}}, \bibinfo {author} {\bibfnamefont {M.}~\bibnamefont
  {Človečko}}, \bibinfo {author} {\bibfnamefont {V.~B.}\ \bibnamefont
  {Eltsov}}, \bibinfo {author} {\bibfnamefont {R.}~\bibnamefont {de~Graaf}},
  \bibinfo {author} {\bibfnamefont {J.}~\bibnamefont {Hosio}}, \bibinfo
  {author} {\bibfnamefont {M.}~\bibnamefont {Krusius}}, \bibinfo {author}
  {\bibfnamefont {D.}~\bibnamefont {Schmoranzer}}, \bibinfo {author}
  {\bibfnamefont {W.}~\bibnamefont {Schoepe}}, \bibinfo {author} {\bibfnamefont
  {L.}~\bibnamefont {Skrbek}}, \bibinfo {author} {\bibfnamefont
  {P.}~\bibnamefont {Skyba}}, \bibinfo {author} {\bibfnamefont {R.~E.}\
  \bibnamefont {Solntsev}},\ and\ \bibinfo {author} {\bibfnamefont {D.~E.}\
  \bibnamefont {Zmeev}},\ }\bibfield  {title} {\bibinfo {title} {Quartz tuning
  fork: Thermometer, pressure- and viscometer for helium liquids},\ }\href
  {https://doi.org/10.1007/s10909-006-9279-4} {\bibfield  {journal} {\bibinfo
  {journal} {Journal of Low Temperature Physics}\ }\textbf {\bibinfo {volume}
  {146}},\ \bibinfo {pages} {537} (\bibinfo {year} {2007})}\BibitemShut
  {NoStop}%
\bibitem [{\citenamefont {Bradley}\ \emph {et~al.}(2009)\citenamefont
  {Bradley}, \citenamefont {Crookston}, \citenamefont {Fisher}, \citenamefont
  {Ganshin}, \citenamefont {Guénault}, \citenamefont {Haley}, \citenamefont
  {Jackson}, \citenamefont {Pickett}, \citenamefont {Schanen},\ and\
  \citenamefont {Tsepelin}}]{Bradley2009}%
  \BibitemOpen
  \bibfield  {author} {\bibinfo {author} {\bibfnamefont {D.~I.}\ \bibnamefont
  {Bradley}}, \bibinfo {author} {\bibfnamefont {P.}~\bibnamefont {Crookston}},
  \bibinfo {author} {\bibfnamefont {S.~N.}\ \bibnamefont {Fisher}}, \bibinfo
  {author} {\bibfnamefont {A.}~\bibnamefont {Ganshin}}, \bibinfo {author}
  {\bibfnamefont {A.~M.}\ \bibnamefont {Guénault}}, \bibinfo {author}
  {\bibfnamefont {R.~P.}\ \bibnamefont {Haley}}, \bibinfo {author}
  {\bibfnamefont {M.~J.}\ \bibnamefont {Jackson}}, \bibinfo {author}
  {\bibfnamefont {G.~R.}\ \bibnamefont {Pickett}}, \bibinfo {author}
  {\bibfnamefont {R.}~\bibnamefont {Schanen}},\ and\ \bibinfo {author}
  {\bibfnamefont {V.}~\bibnamefont {Tsepelin}},\ }\bibfield  {title} {\bibinfo
  {title} {The damping of a quartz tuning fork in superfluid $^3${He}-{B} at
  low temperatures},\ }\href {https://doi.org/10.1007/s10909-009-9982-z}
  {\bibfield  {journal} {\bibinfo  {journal} {Journal of Low Temperature
  Physics}\ }\textbf {\bibinfo {volume} {157}},\ \bibinfo {pages} {476}
  (\bibinfo {year} {2009})}\BibitemShut {NoStop}%
\bibitem [{\citenamefont {Yano}\ \emph {et~al.}(2005)\citenamefont {Yano},
  \citenamefont {Handa}, \citenamefont {Nakagawa}, \citenamefont {Obara},
  \citenamefont {Ishikawa}, \citenamefont {Hata},\ and\ \citenamefont
  {Nakagawa}}]{Yano2005}%
  \BibitemOpen
  \bibfield  {author} {\bibinfo {author} {\bibfnamefont {H.}~\bibnamefont
  {Yano}}, \bibinfo {author} {\bibfnamefont {A.}~\bibnamefont {Handa}},
  \bibinfo {author} {\bibfnamefont {H.}~\bibnamefont {Nakagawa}}, \bibinfo
  {author} {\bibfnamefont {K.}~\bibnamefont {Obara}}, \bibinfo {author}
  {\bibfnamefont {O.}~\bibnamefont {Ishikawa}}, \bibinfo {author}
  {\bibfnamefont {T.}~\bibnamefont {Hata}},\ and\ \bibinfo {author}
  {\bibfnamefont {M.}~\bibnamefont {Nakagawa}},\ }\bibfield  {title} {\bibinfo
  {title} {Observation of laminar and turbulent flow in superfluid $^4${He}
  using a vibrating wire},\ }\href {https://doi.org/10.1007/s10909-005-2261-8}
  {\bibfield  {journal} {\bibinfo  {journal} {Journal of Low Temperature
  Physics}\ }\textbf {\bibinfo {volume} {138}},\ \bibinfo {pages} {561}
  (\bibinfo {year} {2005})}\BibitemShut {NoStop}%
\bibitem [{\citenamefont {Efimov}\ \emph {et~al.}(2009)\citenamefont {Efimov},
  \citenamefont {Garg}, \citenamefont {Giltrow}, \citenamefont {McClintock},
  \citenamefont {Skrbek},\ and\ \citenamefont {Vinen}}]{Efimov2009}%
  \BibitemOpen
  \bibfield  {author} {\bibinfo {author} {\bibfnamefont {V.~B.}\ \bibnamefont
  {Efimov}}, \bibinfo {author} {\bibfnamefont {D.}~\bibnamefont {Garg}},
  \bibinfo {author} {\bibfnamefont {M.}~\bibnamefont {Giltrow}}, \bibinfo
  {author} {\bibfnamefont {P.~V.~E.}\ \bibnamefont {McClintock}}, \bibinfo
  {author} {\bibfnamefont {L.}~\bibnamefont {Skrbek}},\ and\ \bibinfo {author}
  {\bibfnamefont {W.~F.}\ \bibnamefont {Vinen}},\ }\bibfield  {title} {\bibinfo
  {title} {Experiments on a high quality grid oscillating in superfluid
  $^4${He} at very low temperatures},\ }\href
  {https://doi.org/10.1007/s10909-009-9992-x} {\bibfield  {journal} {\bibinfo
  {journal} {Journal of Low Temperature Physics}\ }\textbf {\bibinfo {volume}
  {158}},\ \bibinfo {pages} {462} (\bibinfo {year} {2009})}\BibitemShut
  {NoStop}%
\bibitem [{\citenamefont {Jäger}\ \emph {et~al.}(1995)\citenamefont {Jäger},
  \citenamefont {Schuderer},\ and\ \citenamefont {Schoepe}}]{Jager1995}%
  \BibitemOpen
  \bibfield  {author} {\bibinfo {author} {\bibfnamefont {J.}~\bibnamefont
  {Jäger}}, \bibinfo {author} {\bibfnamefont {B.}~\bibnamefont {Schuderer}},\
  and\ \bibinfo {author} {\bibfnamefont {W.}~\bibnamefont {Schoepe}},\
  }\bibfield  {title} {\bibinfo {title} {Translational oscillations of a
  microsphere in superfluid helium},\ }\href
  {https://doi.org/https://doi.org/10.1016/0921-4526(94)01108-D} {\bibfield
  {journal} {\bibinfo  {journal} {Physica B: Condensed Matter}\ }\textbf
  {\bibinfo {volume} {210}},\ \bibinfo {pages} {201} (\bibinfo {year}
  {1995})}\BibitemShut {NoStop}%
\bibitem [{\citenamefont {Defoort}\ \emph {et~al.}(2016)\citenamefont
  {Defoort}, \citenamefont {Dufresnes}, \citenamefont {Ahlstrom}, \citenamefont
  {Bradley}, \citenamefont {Haley}, \citenamefont {Guénault}, \citenamefont
  {Guise}, \citenamefont {Pickett}, \citenamefont {Poole}, \citenamefont
  {Woods}, \citenamefont {Tsepelin}, \citenamefont {Fisher}, \citenamefont
  {Godfrin},\ and\ \citenamefont {Collin}}]{Defoort2016}%
  \BibitemOpen
  \bibfield  {author} {\bibinfo {author} {\bibfnamefont {M.}~\bibnamefont
  {Defoort}}, \bibinfo {author} {\bibfnamefont {S.}~\bibnamefont {Dufresnes}},
  \bibinfo {author} {\bibfnamefont {S.~L.}\ \bibnamefont {Ahlstrom}}, \bibinfo
  {author} {\bibfnamefont {D.~I.}\ \bibnamefont {Bradley}}, \bibinfo {author}
  {\bibfnamefont {R.~P.}\ \bibnamefont {Haley}}, \bibinfo {author}
  {\bibfnamefont {A.~M.}\ \bibnamefont {Guénault}}, \bibinfo {author}
  {\bibfnamefont {E.~A.}\ \bibnamefont {Guise}}, \bibinfo {author}
  {\bibfnamefont {G.~R.}\ \bibnamefont {Pickett}}, \bibinfo {author}
  {\bibfnamefont {M.}~\bibnamefont {Poole}}, \bibinfo {author} {\bibfnamefont
  {A.~J.}\ \bibnamefont {Woods}}, \bibinfo {author} {\bibfnamefont
  {V.}~\bibnamefont {Tsepelin}}, \bibinfo {author} {\bibfnamefont {S.~N.}\
  \bibnamefont {Fisher}}, \bibinfo {author} {\bibfnamefont {H.}~\bibnamefont
  {Godfrin}},\ and\ \bibinfo {author} {\bibfnamefont {E.}~\bibnamefont
  {Collin}},\ }\href@noop {} {\bibinfo {title} {Probing {B}ogoliubov
  quasiparticles in superfluid $^3${H}e with a ‘vibrating-wire like’ {MEMS}
  device}} (\bibinfo {year} {2016})\BibitemShut {NoStop}%
\bibitem [{\citenamefont {Gu\'enault}\ \emph {et~al.}(2019)\citenamefont
  {Gu\'enault}, \citenamefont {Guthrie}, \citenamefont {Haley}, \citenamefont
  {Kafanov}, \citenamefont {Pashkin}, \citenamefont {Pickett}, \citenamefont
  {Poole}, \citenamefont {Schanen}, \citenamefont {Tsepelin}, \citenamefont
  {Zmeev}, \citenamefont {Collin}, \citenamefont {Maillet},\ and\ \citenamefont
  {Gazizulin}}]{Guenault2019}%
  \BibitemOpen
  \bibfield  {author} {\bibinfo {author} {\bibfnamefont {A.~M.}\ \bibnamefont
  {Gu\'enault}}, \bibinfo {author} {\bibfnamefont {A.}~\bibnamefont {Guthrie}},
  \bibinfo {author} {\bibfnamefont {R.~P.}\ \bibnamefont {Haley}}, \bibinfo
  {author} {\bibfnamefont {S.}~\bibnamefont {Kafanov}}, \bibinfo {author}
  {\bibfnamefont {Y.~A.}\ \bibnamefont {Pashkin}}, \bibinfo {author}
  {\bibfnamefont {G.~R.}\ \bibnamefont {Pickett}}, \bibinfo {author}
  {\bibfnamefont {M.}~\bibnamefont {Poole}}, \bibinfo {author} {\bibfnamefont
  {R.}~\bibnamefont {Schanen}}, \bibinfo {author} {\bibfnamefont
  {V.}~\bibnamefont {Tsepelin}}, \bibinfo {author} {\bibfnamefont {D.~E.}\
  \bibnamefont {Zmeev}}, \bibinfo {author} {\bibfnamefont {E.}~\bibnamefont
  {Collin}}, \bibinfo {author} {\bibfnamefont {O.}~\bibnamefont {Maillet}},\
  and\ \bibinfo {author} {\bibfnamefont {R.}~\bibnamefont {Gazizulin}},\
  }\bibfield  {title} {\bibinfo {title} {Probing superfluid $^{4}\mathrm{He}$
  with high-frequency nanomechanical resonators down to millikelvin
  temperatures},\ }\href {https://doi.org/10.1103/PhysRevB.100.020506}
  {\bibfield  {journal} {\bibinfo  {journal} {Phys. Rev. B}\ }\textbf {\bibinfo
  {volume} {100}},\ \bibinfo {pages} {020506(R)} (\bibinfo {year}
  {2019})}\BibitemShut {NoStop}%
\bibitem [{\citenamefont {Zheng}\ \emph {et~al.}(2016)\citenamefont {Zheng},
  \citenamefont {Jiang}, \citenamefont {Barquist}, \citenamefont {Lee},\ and\
  \citenamefont {Chan}}]{Zheng2016}%
  \BibitemOpen
  \bibfield  {author} {\bibinfo {author} {\bibfnamefont {P.}~\bibnamefont
  {Zheng}}, \bibinfo {author} {\bibfnamefont {W.~G.}\ \bibnamefont {Jiang}},
  \bibinfo {author} {\bibfnamefont {C.~S.}\ \bibnamefont {Barquist}}, \bibinfo
  {author} {\bibfnamefont {Y.}~\bibnamefont {Lee}},\ and\ \bibinfo {author}
  {\bibfnamefont {H.~B.}\ \bibnamefont {Chan}},\ }\bibfield  {title} {\bibinfo
  {title} {Anomalous damping of a microelectromechanical oscillator in
  superfluid $^{3}\mathrm{He}$-{B}},\ }\href
  {https://doi.org/10.1103/PhysRevLett.117.195301} {\bibfield  {journal}
  {\bibinfo  {journal} {Phys. Rev. Lett.}\ }\textbf {\bibinfo {volume} {117}},\
  \bibinfo {pages} {195301} (\bibinfo {year} {2016})}\BibitemShut {NoStop}%
\bibitem [{\citenamefont {Kamppinen}\ and\ \citenamefont
  {Eltsov}(2019)}]{Kamppinen2019}%
  \BibitemOpen
  \bibfield  {author} {\bibinfo {author} {\bibfnamefont {T.}~\bibnamefont
  {Kamppinen}}\ and\ \bibinfo {author} {\bibfnamefont {V.~B.}\ \bibnamefont
  {Eltsov}},\ }\bibfield  {title} {\bibinfo {title} {Nanomechanical resonators
  for cryogenic research},\ }\href {https://doi.org/10.1007/s10909-018-02124-z}
  {\bibfield  {journal} {\bibinfo  {journal} {Journal of Low Temperature
  Physics}\ }\textbf {\bibinfo {volume} {196}},\ \bibinfo {pages} {283}
  (\bibinfo {year} {2019})}\BibitemShut {NoStop}%
\bibitem [{\citenamefont {Collin}\ \emph {et~al.}(2010)\citenamefont {Collin},
  \citenamefont {Bunkov},\ and\ \citenamefont {Godfrin}}]{Collin2010}%
  \BibitemOpen
  \bibfield  {author} {\bibinfo {author} {\bibfnamefont {E.}~\bibnamefont
  {Collin}}, \bibinfo {author} {\bibfnamefont {Y.~M.}\ \bibnamefont {Bunkov}},\
  and\ \bibinfo {author} {\bibfnamefont {H.}~\bibnamefont {Godfrin}},\
  }\bibfield  {title} {\bibinfo {title} {Addressing geometric nonlinearities
  with cantilever microelectromechanical systems: Beyond the duffing model},\
  }\href {https://doi.org/10.1103/PhysRevB.82.235416} {\bibfield  {journal}
  {\bibinfo  {journal} {Phys. Rev. B}\ }\textbf {\bibinfo {volume} {82}},\
  \bibinfo {pages} {235416} (\bibinfo {year} {2010})}\BibitemShut {NoStop}%
\bibitem [{\citenamefont {Abraham}\ \emph {et~al.}(1970)\citenamefont
  {Abraham}, \citenamefont {Eckstein}, \citenamefont {Ketterson}, \citenamefont
  {Kuchnir},\ and\ \citenamefont {Roach}}]{Abraham1970}%
  \BibitemOpen
  \bibfield  {author} {\bibinfo {author} {\bibfnamefont {B.~M.}\ \bibnamefont
  {Abraham}}, \bibinfo {author} {\bibfnamefont {Y.}~\bibnamefont {Eckstein}},
  \bibinfo {author} {\bibfnamefont {J.~B.}\ \bibnamefont {Ketterson}}, \bibinfo
  {author} {\bibfnamefont {M.}~\bibnamefont {Kuchnir}},\ and\ \bibinfo {author}
  {\bibfnamefont {P.~R.}\ \bibnamefont {Roach}},\ }\bibfield  {title} {\bibinfo
  {title} {Velocity of sound, density, and gr\"uneisen constant in liquid
  $^{4}\mathrm{He}$},\ }\href {https://doi.org/10.1103/PhysRevA.1.250}
  {\bibfield  {journal} {\bibinfo  {journal} {Phys. Rev. A}\ }\textbf {\bibinfo
  {volume} {1}},\ \bibinfo {pages} {250} (\bibinfo {year} {1970})}\BibitemShut
  {NoStop}%
\bibitem [{\citenamefont {Donnelly}\ and\ \citenamefont
  {Barenghi}(1998)}]{Donnelly1998}%
  \BibitemOpen
  \bibfield  {author} {\bibinfo {author} {\bibfnamefont {R.~J.}\ \bibnamefont
  {Donnelly}}\ and\ \bibinfo {author} {\bibfnamefont {C.~F.}\ \bibnamefont
  {Barenghi}},\ }\bibfield  {title} {\bibinfo {title} {The observed properties
  of liquid helium at the saturated vapor pressure},\ }\href
  {https://doi.org/10.1063/1.556028} {\bibfield  {journal} {\bibinfo  {journal}
  {Journal of Physical and Chemical Reference Data}\ }\textbf {\bibinfo
  {volume} {27}},\ \bibinfo {pages} {1217} (\bibinfo {year}
  {1998})}\BibitemShut {NoStop}%
\bibitem [{\citenamefont {Godfrin}\ \emph {et~al.}(2021)\citenamefont
  {Godfrin}, \citenamefont {Beauvois}, \citenamefont {Sultan}, \citenamefont
  {Krotscheck}, \citenamefont {Dawidowski}, \citenamefont {F\aa{}k},\ and\
  \citenamefont {Ollivier}}]{Godfrin2021}%
  \BibitemOpen
  \bibfield  {author} {\bibinfo {author} {\bibfnamefont {H.}~\bibnamefont
  {Godfrin}}, \bibinfo {author} {\bibfnamefont {K.}~\bibnamefont {Beauvois}},
  \bibinfo {author} {\bibfnamefont {A.}~\bibnamefont {Sultan}}, \bibinfo
  {author} {\bibfnamefont {E.}~\bibnamefont {Krotscheck}}, \bibinfo {author}
  {\bibfnamefont {J.}~\bibnamefont {Dawidowski}}, \bibinfo {author}
  {\bibfnamefont {B.}~\bibnamefont {F\aa{}k}},\ and\ \bibinfo {author}
  {\bibfnamefont {J.}~\bibnamefont {Ollivier}},\ }\bibfield  {title} {\bibinfo
  {title} {Dispersion relation of landau elementary excitations and
  thermodynamic properties of superfluid $^{4}\mathrm{He}$},\ }\href
  {https://doi.org/10.1103/PhysRevB.103.104516} {\bibfield  {journal} {\bibinfo
   {journal} {Phys. Rev. B}\ }\textbf {\bibinfo {volume} {103}},\ \bibinfo
  {pages} {104516} (\bibinfo {year} {2021})}\BibitemShut {NoStop}%
\bibitem [{\citenamefont {Brooks}\ and\ \citenamefont
  {Donnelly}(1977)}]{Brooks1977}%
  \BibitemOpen
  \bibfield  {author} {\bibinfo {author} {\bibfnamefont {J.~S.}\ \bibnamefont
  {Brooks}}\ and\ \bibinfo {author} {\bibfnamefont {R.~J.}\ \bibnamefont
  {Donnelly}},\ }\bibfield  {title} {\bibinfo {title} {The calculated
  thermodynamic properties of superfluid helium‐4},\ }\href
  {https://doi.org/10.1063/1.555549} {\bibfield  {journal} {\bibinfo  {journal}
  {Journal of Physical and Chemical Reference Data}\ }\textbf {\bibinfo
  {volume} {6}},\ \bibinfo {pages} {51} (\bibinfo {year} {1977})}\BibitemShut
  {NoStop}%
\bibitem [{\citenamefont {Mc~Carty}(1973)}]{Carty1973}%
  \BibitemOpen
  \bibfield  {author} {\bibinfo {author} {\bibfnamefont {R.~D.}\ \bibnamefont
  {Mc~Carty}},\ }\bibfield  {title} {\bibinfo {title} {Thermodynamic properties
  of helium 4 from 2 to 1500 {K} at pressures to $10^8$ {Pa}},\ }\href
  {https://doi.org/10.1063/1.3253133} {\bibfield  {journal} {\bibinfo
  {journal} {Journal of Physical and Chemical Reference Data}\ }\textbf
  {\bibinfo {volume} {2}},\ \bibinfo {pages} {923} (\bibinfo {year}
  {1973})}\BibitemShut {NoStop}%
\bibitem [{\citenamefont {Sader}(1998)}]{Sader1998}%
  \BibitemOpen
  \bibfield  {author} {\bibinfo {author} {\bibfnamefont {J.~E.}\ \bibnamefont
  {Sader}},\ }\bibfield  {title} {\bibinfo {title} {Frequency response of
  cantilever beams immersed in viscous fluids with applications to the atomic
  force microscope},\ }\href {https://doi.org/10.1063/1.368002} {\bibfield
  {journal} {\bibinfo  {journal} {Journal of Applied Physics}\ }\textbf
  {\bibinfo {volume} {84}},\ \bibinfo {pages} {64} (\bibinfo {year}
  {1998})}\BibitemShut {NoStop}%
\bibitem [{\citenamefont {Fear}\ \emph {et~al.}(2016)\citenamefont {Fear},
  \citenamefont {Walmsley}, \citenamefont {Zmeev}, \citenamefont {Mäkinen},\
  and\ \citenamefont {Golov}}]{Fear2016}%
  \BibitemOpen
  \bibfield  {author} {\bibinfo {author} {\bibfnamefont {M.~J.}\ \bibnamefont
  {Fear}}, \bibinfo {author} {\bibfnamefont {P.~M.}\ \bibnamefont {Walmsley}},
  \bibinfo {author} {\bibfnamefont {D.~E.}\ \bibnamefont {Zmeev}}, \bibinfo
  {author} {\bibfnamefont {J.~T.}\ \bibnamefont {Mäkinen}},\ and\ \bibinfo
  {author} {\bibfnamefont {A.~I.}\ \bibnamefont {Golov}},\ }\bibfield  {title}
  {\bibinfo {title} {No effect of steady rotation on solid $^4${He} in a
  torsional oscillator},\ }\href {https://doi.org/10.1007/s10909-015-1376-9}
  {\bibfield  {journal} {\bibinfo  {journal} {Journal of Low Temperature
  Physics}\ }\textbf {\bibinfo {volume} {183}},\ \bibinfo {pages} {106}
  (\bibinfo {year} {2016})}\BibitemShut {NoStop}%
\bibitem [{\citenamefont {Vinen}\ and\ \citenamefont
  {Skrbek}(2014)}]{Vinen2014}%
  \BibitemOpen
  \bibfield  {author} {\bibinfo {author} {\bibfnamefont {W.~F.}\ \bibnamefont
  {Vinen}}\ and\ \bibinfo {author} {\bibfnamefont {L.}~\bibnamefont {Skrbek}},\
  }\bibfield  {title} {\bibinfo {title} {Quantum turbulence generated by
  oscillating structures},\ }\href {https://doi.org/10.1073/pnas.1312551111}
  {\bibfield  {journal} {\bibinfo  {journal} {Proceedings of the National
  Academy of Sciences}\ }\textbf {\bibinfo {volume} {111}},\ \bibinfo {pages}
  {4699} (\bibinfo {year} {2014})}\BibitemShut {NoStop}%
\bibitem [{\citenamefont {Phillips}(1987)}]{Phillips1987}%
  \BibitemOpen
  \bibfield  {author} {\bibinfo {author} {\bibfnamefont {W.~A.}\ \bibnamefont
  {Phillips}},\ }\bibfield  {title} {\bibinfo {title} {Two-level states in
  glasses},\ }\href {http://stacks.iop.org/0034-4885/50/i=12/a=003} {\bibfield
  {journal} {\bibinfo  {journal} {Rep. Prog. Phys.}\ }\textbf {\bibinfo
  {volume} {50}},\ \bibinfo {pages} {1657} (\bibinfo {year}
  {1987})}\BibitemShut {NoStop}%
\bibitem [{\citenamefont {Baym}\ \emph {et~al.}(1969)\citenamefont {Baym},
  \citenamefont {Barrera},\ and\ \citenamefont {Pethick}}]{Baym1969}%
  \BibitemOpen
  \bibfield  {author} {\bibinfo {author} {\bibfnamefont {G.}~\bibnamefont
  {Baym}}, \bibinfo {author} {\bibfnamefont {R.~G.}\ \bibnamefont {Barrera}},\
  and\ \bibinfo {author} {\bibfnamefont {C.~J.}\ \bibnamefont {Pethick}},\
  }\bibfield  {title} {\bibinfo {title} {Mobility of the electron bubble in
  superfluid helium},\ }\href {https://doi.org/10.1103/PhysRevLett.22.20}
  {\bibfield  {journal} {\bibinfo  {journal} {Phys. Rev. Lett.}\ }\textbf
  {\bibinfo {volume} {22}},\ \bibinfo {pages} {20} (\bibinfo {year}
  {1969})}\BibitemShut {NoStop}%
\bibitem [{\citenamefont {Niemetz}\ and\ \citenamefont
  {Schoepe}(2004)}]{Niemetz2004}%
  \BibitemOpen
  \bibfield  {author} {\bibinfo {author} {\bibfnamefont {M.}~\bibnamefont
  {Niemetz}}\ and\ \bibinfo {author} {\bibfnamefont {W.}~\bibnamefont
  {Schoepe}},\ }\bibfield  {title} {\bibinfo {title} {Stability of laminar and
  turbulent flow of superfluid $ ^4 \mathrm{He} $ at {mK} temperatures around
  an oscillating microsphere},\ }\href
  {https://doi.org/10.1023/B:JOLT.0000029507.98543.1d} {\bibfield  {journal}
  {\bibinfo  {journal} {Journal of Low Temperature Physics}\ }\textbf {\bibinfo
  {volume} {135}},\ \bibinfo {pages} {447} (\bibinfo {year}
  {2004})}\BibitemShut {NoStop}%
\bibitem [{\citenamefont {Rayfield}\ and\ \citenamefont
  {Reif}(1964)}]{Rayfield1964}%
  \BibitemOpen
  \bibfield  {author} {\bibinfo {author} {\bibfnamefont {G.~W.}\ \bibnamefont
  {Rayfield}}\ and\ \bibinfo {author} {\bibfnamefont {F.}~\bibnamefont
  {Reif}},\ }\bibfield  {title} {\bibinfo {title} {Quantized vortex rings in
  superfluid helium},\ }\href {https://doi.org/10.1103/PhysRev.136.A1194}
  {\bibfield  {journal} {\bibinfo  {journal} {Phys. Rev.}\ }\textbf {\bibinfo
  {volume} {136}},\ \bibinfo {pages} {A1194} (\bibinfo {year}
  {1964})}\BibitemShut {NoStop}%
\bibitem [{\citenamefont {Yorozu}\ \emph {et~al.}(1993)\citenamefont {Yorozu},
  \citenamefont {Fukuyama},\ and\ \citenamefont {Ishimoto}}]{Yorozu1993}%
  \BibitemOpen
  \bibfield  {author} {\bibinfo {author} {\bibfnamefont {S.}~\bibnamefont
  {Yorozu}}, \bibinfo {author} {\bibfnamefont {H.}~\bibnamefont {Fukuyama}},\
  and\ \bibinfo {author} {\bibfnamefont {H.}~\bibnamefont {Ishimoto}},\
  }\bibfield  {title} {\bibinfo {title} {Isochoric pressure and $^{3}${He}
  quasiparticle effective mass in a $^{3}\mathrm{He}{\ensuremath{-}}^{4}
  \mathrm{He} $ mixture under pressure},\ }\href
  {https://doi.org/10.1103/PhysRevB.48.9660} {\bibfield  {journal} {\bibinfo
  {journal} {Phys. Rev. B}\ }\textbf {\bibinfo {volume} {48}},\ \bibinfo
  {pages} {9660} (\bibinfo {year} {1993})}\BibitemShut {NoStop}%
\bibitem [{\citenamefont {Wakita}\ and\ \citenamefont
  {Sano}(1983)}]{Wakita1983}%
  \BibitemOpen
  \bibfield  {author} {\bibinfo {author} {\bibfnamefont {H.}~\bibnamefont
  {Wakita}}\ and\ \bibinfo {author} {\bibfnamefont {Y.}~\bibnamefont {Sano}},\
  }\bibfield  {title} {\bibinfo {title} {$^3${He}/$^4${He} ratios in {CH4}-rich
  natural gases suggest magmatic origin},\ }\href
  {https://doi.org/10.1038/305792a0} {\bibfield  {journal} {\bibinfo  {journal}
  {Nature}\ }\textbf {\bibinfo {volume} {305}},\ \bibinfo {pages} {792}
  (\bibinfo {year} {1983})}\BibitemShut {NoStop}%
\bibitem [{\citenamefont {Landau}\ and\ \citenamefont
  {Lifshitz}(1987)}]{Landau1987}%
  \BibitemOpen
  \bibfield  {author} {\bibinfo {author} {\bibfnamefont {L.~D.}\ \bibnamefont
  {Landau}}\ and\ \bibinfo {author} {\bibfnamefont {E.~M.}\ \bibnamefont
  {Lifshitz}},\ }\href@noop {} {\emph {\bibinfo {title} {Fluid Mechanics}}},\
  \bibinfo {edition} {2nd}\ ed.\ (\bibinfo  {publisher} {Pergamon Press},\
  \bibinfo {year} {1987})\BibitemShut {NoStop}%
\bibitem [{\citenamefont {Vinen}\ \emph {et~al.}(2003)\citenamefont {Vinen},
  \citenamefont {Tsubota},\ and\ \citenamefont {Mitani}}]{Vinen2003}%
  \BibitemOpen
  \bibfield  {author} {\bibinfo {author} {\bibfnamefont {W.~F.}\ \bibnamefont
  {Vinen}}, \bibinfo {author} {\bibfnamefont {M.}~\bibnamefont {Tsubota}},\
  and\ \bibinfo {author} {\bibfnamefont {A.}~\bibnamefont {Mitani}},\
  }\bibfield  {title} {\bibinfo {title} {Kelvin-wave cascade on a vortex in
  superfluid $^{4}\mathrm{H}\mathrm{e}$ at a very low temperature},\ }\href
  {https://doi.org/10.1103/PhysRevLett.91.135301} {\bibfield  {journal}
  {\bibinfo  {journal} {Phys. Rev. Lett.}\ }\textbf {\bibinfo {volume} {91}},\
  \bibinfo {pages} {135301} (\bibinfo {year} {2003})}\BibitemShut {NoStop}%
\bibitem [{\citenamefont {Eltsov}\ and\ \citenamefont
  {L’vov}(2020)}]{Eltsov2020}%
  \BibitemOpen
  \bibfield  {author} {\bibinfo {author} {\bibfnamefont {V.~B.}\ \bibnamefont
  {Eltsov}}\ and\ \bibinfo {author} {\bibfnamefont {V.~S.}\ \bibnamefont
  {L’vov}},\ }\bibfield  {title} {\bibinfo {title} {Amplitude of waves in the
  kelvin-wave cascade},\ }\href {https://doi.org/10.1134/S0021364020070012}
  {\bibfield  {journal} {\bibinfo  {journal} {JETP Letters}\ }\textbf {\bibinfo
  {volume} {111}},\ \bibinfo {pages} {389} (\bibinfo {year}
  {2020})}\BibitemShut {NoStop}%
\bibitem [{\citenamefont {Shkarin}\ \emph {et~al.}(2019)\citenamefont
  {Shkarin}, \citenamefont {Kashkanova}, \citenamefont {Brown}, \citenamefont
  {Garcia}, \citenamefont {Ott}, \citenamefont {Reichel},\ and\ \citenamefont
  {Harris}}]{Shkarin2019}%
  \BibitemOpen
  \bibfield  {author} {\bibinfo {author} {\bibfnamefont {A.~B.}\ \bibnamefont
  {Shkarin}}, \bibinfo {author} {\bibfnamefont {A.~D.}\ \bibnamefont
  {Kashkanova}}, \bibinfo {author} {\bibfnamefont {C.~D.}\ \bibnamefont
  {Brown}}, \bibinfo {author} {\bibfnamefont {S.}~\bibnamefont {Garcia}},
  \bibinfo {author} {\bibfnamefont {K.}~\bibnamefont {Ott}}, \bibinfo {author}
  {\bibfnamefont {J.}~\bibnamefont {Reichel}},\ and\ \bibinfo {author}
  {\bibfnamefont {J.~G.~E.}\ \bibnamefont {Harris}},\ }\bibfield  {title}
  {\bibinfo {title} {Quantum optomechanics in a liquid},\ }\href
  {https://doi.org/10.1103/PhysRevLett.122.153601} {\bibfield  {journal}
  {\bibinfo  {journal} {Phys. Rev. Lett.}\ }\textbf {\bibinfo {volume} {122}},\
  \bibinfo {pages} {153601} (\bibinfo {year} {2019})}\BibitemShut {NoStop}%
\bibitem [{\citenamefont {Landau}\ \emph {et~al.}(1986)\citenamefont {Landau},
  \citenamefont {Lifshitz}, \citenamefont {Kosevich}, \citenamefont
  {Pitaevskii}, \citenamefont {Sykes},\ and\ \citenamefont
  {Reid}}]{LandauLifshitz1986}%
  \BibitemOpen
  \bibfield  {author} {\bibinfo {author} {\bibfnamefont {L.~D.}\ \bibnamefont
  {Landau}}, \bibinfo {author} {\bibfnamefont {E.~M.}\ \bibnamefont
  {Lifshitz}}, \bibinfo {author} {\bibfnamefont {A.~M.}\ \bibnamefont
  {Kosevich}}, \bibinfo {author} {\bibfnamefont {L.~P.}\ \bibnamefont
  {Pitaevskii}}, \bibinfo {author} {\bibfnamefont {J.~B.}\ \bibnamefont
  {Sykes}},\ and\ \bibinfo {author} {\bibfnamefont {W.~H.}\ \bibnamefont
  {Reid}},\ }\href@noop {} {\emph {\bibinfo {title} {Landau and Lifshitz Course
  of Theoretical Physics, Theory of Elasticity}}},\ \bibinfo {edition} {3rd}\
  ed.,\ Vol.~\bibinfo {volume} {7}\ (\bibinfo  {publisher} {Elsevier},\
  \bibinfo {year} {1986})\BibitemShut {NoStop}%
\end{thebibliography}%


%

\end{document}